%% file: main.tex
\documentclass[10pt, twosided]{article}
\usepackage[utf8]{inputenc}
\usepackage{graphicx}
\graphicspath{{figs/}}
\usepackage{vmargin}
\usepackage{setspace}
\usepackage{amsmath}
\usepackage{amssymb}
\usepackage{url}
\usepackage{pgfplots}
\pgfplotsset{compat=newest}
\usetikzlibrary{patterns}
\usetikzlibrary{calc}
\usepackage{subfigure}
\usepackage{color, colortbl}

\definecolor{Gray}{gray}{0.9}

\newcommand{\R}{\mathbb{R}}

\newcommand{\bs}{\boldsymbol}

\setmarginsrb{2.5cm}{0.8 cm}{2.5 cm}{1.0 cm}{1 cm}{1.5 cm}{1 cm}{1.5 cm}
\title{Visualizing and Understanding  Large-Scale Assessments in Mathematics through Dimensionality Reduction}

\makeatletter
\renewcommand\@date{{%
  \vspace{-\baselineskip}%
  \large\centering
  \begin{tabular}{@{}c@{}}
    Esdras Medeiros\textsuperscript{1} \\
    \footnotesize esdras@mat.ufc.br
  \end{tabular}%
  \begin{tabular}{@{}c@{}}
    Jorge Lira\textsuperscript{1} \\
    \footnotesize jorge.lira@mat.ufc.br
  \end{tabular}
 \begin{tabular}{@{}c@{}}
    Romildo Silva\textsuperscript{2} \\
    \footnotesize rjs@mat.ufc.br
  \end{tabular}
 \begin{tabular}{@{}c@{}}
    Caio Azevedo\textsuperscript{2} \\
    \footnotesize cnaber@ime.unicamp.br
  \end{tabular}

  \bigskip

  {\small \textsuperscript{1}Department of Mathematics, Universidade Federal do
  Cear\'a, Cear\'a, Brazil}\\
  {\small \textsuperscript{2}IME -- Universidade Estadual de Campinas,
  S\~{a}o  Paulo, Brazil}

  \bigskip

  \today
}}
\makeatother

\begin{document}

\maketitle

\begin{abstract}
    In this paper, we apply the Logistic PCA (LPCA)  as a dimensionality reduction tool 
    for visualizing patterns and  characterizing the relevance of mathematics abilities 
    from a given population measured by a large-scale assessment. We
    establish an equivalence of parameters between LPCA, 
    Inner Product Representation (IPR) and the two paramenter logistic model
    (2PL) from the Item Response Theory (IRT). This equivalence provides three
    complemetary ways of looking at data that assists professionals in
    education to  perform in-context interpretations. Particularly,  we analyse 
    the data collected from SPAECE,  a large-scale assessment in Mathematics that has been 
    applied yearly in the public  educational system of the state of
    Cear\'a, Brazil. As the main result, we show that the the poor
    performance of examinees in the end of middle school is primarily caused by
    their disabilities in number sense.  
    
\end{abstract}

\input{introduction.tex}

\input{relatedWork}
\input{background.tex}

\input{applications.tex}
\input{discussion.tex}
\input{acknowledments.tex}

\bibliographystyle{plain}
\bibliography{main}

\appendix

\input{normalDeviance.tex}
\input{bernoulliDiviance.tex}

\input{referenceMatrix}

\end{document}

%% file: introduction.tex
\section{Introduction}

We are witnessing the birth of the fourth industrial revolution, being a technological trend that is transforming the way we live, work and
interact to each other. This new era is driven by a set of technological
innovations such as Robotics, Artificial Intelligence, Big Data (massive
volume data analysis), 3D Printing, Augmented Reality, Synthetic Biology,
Nanotechnology and the commonly known Internet of Things whereby, increasingly
the devices are connected to others through the Internet. Some of which
said innovations are still in their embryonic stage, however being  ready to develop quickly. 

The World Economic Forum held in Davos, Switzerland in 2016,  
published a report~\cite{wefReport2016} which places  the ability of
\textit{complex problem solving} as the most important  skill required for
students to achieve a successful career demanded by this technological
revolution. It has become a consensus that the major mean of developing
problem-solving skills is through Mathematics (see~\cite{nctm2000}~p.52).
Indeed, Mathematics helps us to analyze and think logically about new situations, devise unspecified solution procedures,
and communicate their solution clearly and convincingly to others. 
One of the main goals of learning Mathematics is to deal with abstractions that sometimes model and solve concrete problems 
which are apparently disconnected.


Public Educational Systems need to be aware of the technological advances in
the world and properly prepare students for the job market. However, it has
been a great challenge for (municipal, state, and national) governments to
recycle teachers and improve the  infrastructure and administration of
schools. To assist policy makers in their decisions, large-scale assessments
of Mathematics are used to monitor the abilities and knowledge of students~\cite{Suurtamm2016}. 

SPAECE\footnote{Here we will keep the acronym spelling in portuguese.}
(Permanent Evaluation System of Basic Education in the State of  Cear\'a) is
local large-scale assessment in Portuguese and Mathematics in the state
of Cear\'a. Using IRT, this assessment yearly
collects the proficiency  in these areas from all students attending the public schools of the state. 
This database has been the main reference source to diagnose school results and accountability to society 
by providing the big picture of the quality of public education of Cear\'a.

SPAECE is taken at three levels: at the end of elementary school~(L1), at the end of middle school~(L2) and
at the end of high school~(L3).  SPAECE uses a Reference Matrix (RM)  which is composed of a set of descriptors that
explain the level of mental operation required to perform certain tasks. These descriptors are selected considering what
can be evaluated by means of a multiple choice test, whose items imply the selection of a response in a given set of 
possible answers. There are three RMs, one for each level, and they may share a few descriptors. 
In Appendix~\ref{ap:referenceMatrix} we show the RM corresponding to L3.


Our goal is to provide {\bf mathematical and visual insights} of large-scale
assessments in Mathematics through the LPCA~\cite{2015arXiv151006112L}, a Principal Component Analysis (PCA) tool 
for binary data which has become a popular alternative to dimensionality reduction of binary data. 
Variables  correspond to descriptors of the RM. By applying LPCA to them,
we obtain a set of 1--2 principal components which carry precious
quantitative/qualitative information of the students' proficiency.
By using them,  managers  can evaluate the efficiency 
of the education system as a whole. In addition, as described in
Section~\ref{sec:applications},  education analysts can find insightful patterns
which guide the development of pedagogical actions.

This paper is organized as follows. In Section~\ref{sec:relatedWork} we
discuss related work. In Section~\ref{sec:backgroundPCA} we set up the
mathematical background. More specifically, we establish a connection
between  LPCA, IRT and IPR.  In
Section~\ref{sec:applications} we describe SPAECE data set. Then, we apply
the LPCA  and provide visualization tools. From the patterns we find in the
visualizations, we extract useful information that can  guide improvements
to the educational system of Cear\'a.  Finally, in
Section~\ref{sec:discussion},   we summarize  our contributions and discuss limitations.

%% file: relatedWork.tex
\section{Related work}
\label{sec:relatedWork}

IRT\cite{Reckase2009}, in its many forms, are the most prevalent used models in
large-scale assessment programs. The main feature of IRT is the estimation 
of item difficulties and  examinee abilities separately, but on the same
scale. Furthermore, it can place different tests on the same scale (e.g. linking
and equating). An important property of IRT is the local independence, i.e.,
for examinee at a given location on the scale, the success outcome of any
item does not depend on any other item of that scale.

We can find a vast literature ralated to IRT and applications
because it has been a well developed research area in the last
decades. However, as  far as we know, there are only a few works that explore  visual
aspects of data. For instance, Graphical Item Analysis (GIA) \cite{laros2002} is a method that visually displays the
relationship between the total score on a test and the response proportions
of correct and false alternatives. The main goal is to obtain a visual understanding of IRT and its
item's statistics (e.g. distractors and correct answer
proportions).
It has been used to  
analyse the quality of items of large-scale assessments of Portuguese and
Mathematics in Brazil. 
A similar approach was also proposed in~\cite{livingston2004} for tests of
proficiency in English.  
While such works concentrate on 
improving the quality of the assessment \textit{per se}, our work aims at providing 
visualizations to understand the  entire proficiency of the population.

According to Colin Ware~\cite{Ware12}, the term {\it Visualization} means  something more like a graphical representation 
of data or concepts. In a nutshell, visualization  combines human factors and data analysis to gain insight in the problem 
at hand~\cite{cook2015}.  Our contribution goes in this direction, more specifically,  
we do \textit{Visual Analytics}~\cite{cook2015, keim2006} using generalized PCA as a
dimensionality reduction technique. From the results delivered by  this
technique, we generate plots  that make easier human perception of patterns. In the context of large-scale
asessments, patterns are translated to useful information that captures the actual status of educational systems.

PCA is a dimensionality reduction method primarily designed for real-valued data 
that minimizes a squared loss function~\cite{Pearson1901}.   It can be formulated as a maximum
likelihood estimation (MLE) problem~\cite{Tipping99}. Such probabilistic
view can be extended, in a natural way, to the exponential family of
distributions~\cite{dasgupta2002, deLeeuw2006,
2015arXiv151006112L}.  The LPCA is one of such extensions that we apply to the dichotomized
responses of items  which are correct--1 and wrong--0. 

There are equivalences between dimensionality reduction methods and IRT.
In \cite{leeuw1987},  it is shown a relationship between 
binary variables in Factor Analysis (FA)~\cite{Thurstone1936, Child1990} and
the two-parameter normal ogive model in IRT model. In~\cite{groenen2003}, 
it is presented a connection  beween the bi-additive
logistic model~\cite{Denis1994} and the two-parameter logistic IRT model.
It is well known that FA and bi-additive models are related to PCA but they
are conceptually disticts and the results may not be
identical~\cite{SANTOS2019, Gower2006}.
In this paper, we describe  connections between  LPCA and IPR with
two-parameter IRT model through the equivalence of parameters
as shown in Table~\ref{tb:equivalence}. This equivalence may help to better understand the
proposed visualization tools in Section~\ref{sec:applications}. 

%% file: background.tex
\section{Background}
\label{sec:backgroundPCA}
 
\subsection{Conventional PCA}

Assume that it is given a matrix $X_{n\times d}$ where every row  is a vector  
$\bs{x_i} \in \R^d$, $i=1,2...,n$, representing a set of observations and each column corresponds to an observable variable. 
For $k<d$, PCA is a technique that attempts to find a $k$-dimensional subspace passing
close to the data. One way to solve this problem is originally proposed in~\cite{Pearson1901}.
The key idea is to find a coordinate system, i.e., a translation vector $\bs{\mu} \in \R^d$ and a
set vectors (also called \emph{principal components}) $\bs{u_1},
\bs{u_2},..., \bs{u_k} \in \R^d$ , forming a matrix $U_{d\times k } =
[\bs{u_1}\; \bs{u_2}\;...\bs{u_k}]$, such that the projected data
$\displaystyle \bs{\theta_i}^T
= \bs{\mu} + UU^T(\bs{x_i}^T-\bs{\mu})$  is ``close" to $\bs{x_i}^T$ as
possible and which belong to a $k$-dimensional subspace. More
specifically,  in a conventional PCA, the goal is to minimize the sum of squared distances from data points
$\bs{x_i}$ to their projections $\bs{\theta_i}$, that is,

\begin{equation}
\label{eq:obj_func}
\min \sum_{i=1}^n||\bs{x_i} - \bs{\theta_i}||^2 = \min \sum_{i=1}^n||(\bs{x_i}-\bs{\mu}) - UU^T(\bs{x_i}-\bs{\mu})||^2
\end{equation}

\noindent over $\bs{\mu}$ and $d\times k$ orthogonal matrix $U$.
 We denominate each coordinate of $\bs{u_i}$ as  {\it loading}.

\subsection{Generalized PCA to Exponential Family}

The objective function expressed in (\ref{eq:obj_func}),   
can alternatively be derived from a probabilistic
perspective~\cite{Tipping99} which allows a natural  generalization of
conventional PCA to exponential 
family distributions~\cite{dasgupta2002,deLeeuw2006, 2015arXiv151006112L}.
In the exponential family the conditional probability of  $x$, given \textit{natural
parameter} $\theta$, has the following canonical form (or natural form):

\begin{equation}
\label{eq:exp_family}
    \log f(x ; \theta) = x\theta - b(\theta) + c(x)
\end{equation}

\noindent where $E[x\,;\, \theta ] = b'(\theta)$  and $c(\cdot)$ 
is a function that depends only on $x$ and can be discarded during optimization process.
The function  $g(\cdot)$  such that $g(b'(\theta))=\theta$ is called the
\textit{canonical link function}. Any distribution  with parameter $\alpha$ can be converted to a
canonical form by  appyling the transformation~$g(\alpha) = \theta$. If $g =
\mathbf{id}$, where $\mathbf{id}$ is the indentity function,  the
distribution is already in the canonical form.

In a normal distribution, with mean $\mu$ and unit variance, the
density is usually described  as $\log f(x  ;  \phi) = -\log{\sqrt{2\pi}}
- (x-\phi)^2/2$. This distribution belongs to the explonential family  with
$c(x)=  -\log{\sqrt{2\pi}} + x^2/2$, $g = \mathbf{id}$ and
$b(\theta)=\theta^2/2$.  Note
that the normal distribution is already in the canonical form because $g$ is
the identity function. Another common example is the Bernoulli distribution
for binary outcomes $x \in \{0, 1\}$. The probability is written as $f(x 
 ; p) = p^x(1-p)^{(1-x)}$,  where $p$ is a parameter in $[0, 1]$. In this
distribution we  get $c(x) = 0$, $g = \textrm{logit}$ and $b(\theta) =
\log(1+ e^\theta)$.


Let $X$ be the matrix of observed data with distributions having parameters
represented by the  matrix~$\Theta$. The $\log$ likelihood expressed as a function of $\Theta$ is: 

\begin{equation}
\label{eq:likelihood}
l(\Theta;X) = \sum_{i, j} \log \; f(x_{ij}; \theta_{ij})
\end{equation}

A  common goodness-of-fit for estimation is given by the
\textit{scaled deviance}\cite{McCullagh1989}. The
scaled deviance $D(X, \Theta)$ express how the likelihood   
$l(\Theta, X)$  diverges from the likelihood of 
the \textit{saturated model} (the maximum likelihood attainable for an exact fit in
which the fitted values are equal to the observed data) $l(X, X)$ and is  given by:

\begin{equation}
\label{eq:scaledDeviation}
D(X;\Theta) = 2l(X, X) - 2l(\Theta, X)
\end{equation}



The objective function for PCA to exponential family proposed in $\cite{2015arXiv151006112L}$ 
minimizes the deviance of projection of the natural parameters from the
saturated model. More specifically, let $\tilde{\theta}=g(x)$ be the natural 
parameter for the saturated model  and 

\begin{equation}
\label{eq:projection}
\bs{\theta_i}^T = \bs{\mu} +
UU^T(\tilde{\bs{\theta}}_{\bs{i}}^T-\bs{\mu})
\end{equation}

\noindent  be its projection to the $k$-dimensional space. The generalized
PCA optimization problem corresponds to the minimization of the deviance
$D(X, \Theta)$ with respect to $U$ and $\mu$, which can be cast as:

\begin{align}
\min D(X, \Theta)  & = \min 2l(X, X)-2l(\Theta, X) =  \min - l(\Theta, X) =
    -\sum_{i, j}\log f(x_{ij}\; ;\theta_{ij}) =  \nonumber \\
                  & = \min   \sum_{i, j} -x_{ij}\theta_{ij} + b(\theta_{ij})
                  = \min -tr(X\Theta^T) + \sum_{i, j} b(\theta_{ij}) =
                  \nonumber \\ 
                  & =  \min -tr\left( X(\bs{\mu}(\bs{1_n})^T +
                      (\tilde{\Theta}^T-\bs{\mu}(\bs{1_n})^T)UU^T)\right)  +
                        \sum_{i, j} b(\mu_j + [UU^T(\tilde{\bs{\theta}}_{\bs{i}} - \mu)]_j)
\end{align}

When $x_{ij}$ is an observation from normal distribution
given~by ~$x_{ij} \sim N(\phi_{ij}, 1)$, the natural parameter is
$\theta_{ij}= \phi_{ij}$ and the  
natural parameter from the saturated model is $x_{ij}$ itself.  By
minimizing the deviance~(\ref{eq:scaledDeviation}) for this distribution,  we
obtain exactly the Pearson's  formulation (\ref{eq:obj_func})  (see Appendix~\ref{ap:normalDeviance} for details).


\subsection{Using Logistic PCA for Assessments}
\label{sub:LPCA}

Let  $X$ be a binary data matrix where $x_{ij}$ is equal to zero or one, representing a correct/incorrect item responded by a student. 
Each line correspond to a student and each column correspond to an item.

Let  $P =\{p_{ij}\}$ be a matrix where  $p_{ij}$ is the probability of student $i$ answer correctly an item $j$, 
which  means that $p_{ij} = f(x_{ij}\; ; p_{ij})$,  
where  $f(\cdot\; ;p_{ij})$ is a Bernoulli mass function  with  parameter~$p_{ij}$.
The canonical link function  for the Bernoulli 
distribution provides the natural parameter given by 

\begin{equation}
\label{eq:link_function}
\theta_{ij} = \textrm{logit}\;p_{ij}.
\end{equation}
    As detailed in
Appendix~\ref{ap:bernoulliDeviance},  the scaled deviance of this
distribution  is:


\begin{equation}
\label{eq:bernoulli_deviance}
    D(X, \Theta) = - 2tr(X \Theta^T) + 2\sum_{ij} \log (1+\exp\,\theta_{ij})
\end{equation}

\noindent and  each element of matrix $\Theta = \{\theta_{ij}\}$ varies from
$-\infty$ to $+\infty$. Let $\tilde{\Theta} = \{\tilde{\theta}_{ij}\}$ 
represent the natural parameter for the saturated model. Notice that
$\tilde{\theta}_{ij}$ is $\infty$ 
if $x_{ij}=1$ and $-\infty$ if $x_{ij}=0$, which is unfeasible for numerical computations. 

To extend principal component analysis to binary data and make it computationally realizable, 
it is necessary to limit this domain. For convenience, 
first define $q_{ij} = 2x_{ij} - 1$, which converts the 
binary variable from taking 
values in $\{0, 1\}$ to $\{-1, 1\}$. Let $Q$
be the matrix with elements $q_{ij}$. 
\cite{2015arXiv151006112L} suggest that one can approximate
$\tilde{\theta}_{ij}$ by 
$m\cdot q_{ij}$  for a large number $m$. 
Therefore, $\tilde{\Theta}$ can be approximated by the matrix $m\cdot Q$. As
in conventional PCA, for $k<d$, 
we seek for  vector $\bs{\mu} \in \R^d$  and a matrix $U_{d\times k}$  that
minimizes the Bernoulli deviance:

\begin{equation}
\label{eq:approximation}
    \min  D(X, \Theta) = \min - 2tr(X \Theta^T) + 2\sum_{ij} \log (1+\exp( \mu_j + [UU^T(\tilde{\bs{\theta}}_{\bs{i}} - \mu)]_j)
\end{equation}


The numerical optimization  has been  implemented through an R \cite{R2013} 
public  package available in the web. 
For the web address and technical details we refer the reader to  \cite{2015arXiv151006112L}. 

\begin{figure}[t]
    \centering
 \includegraphics[height=7.5cm]{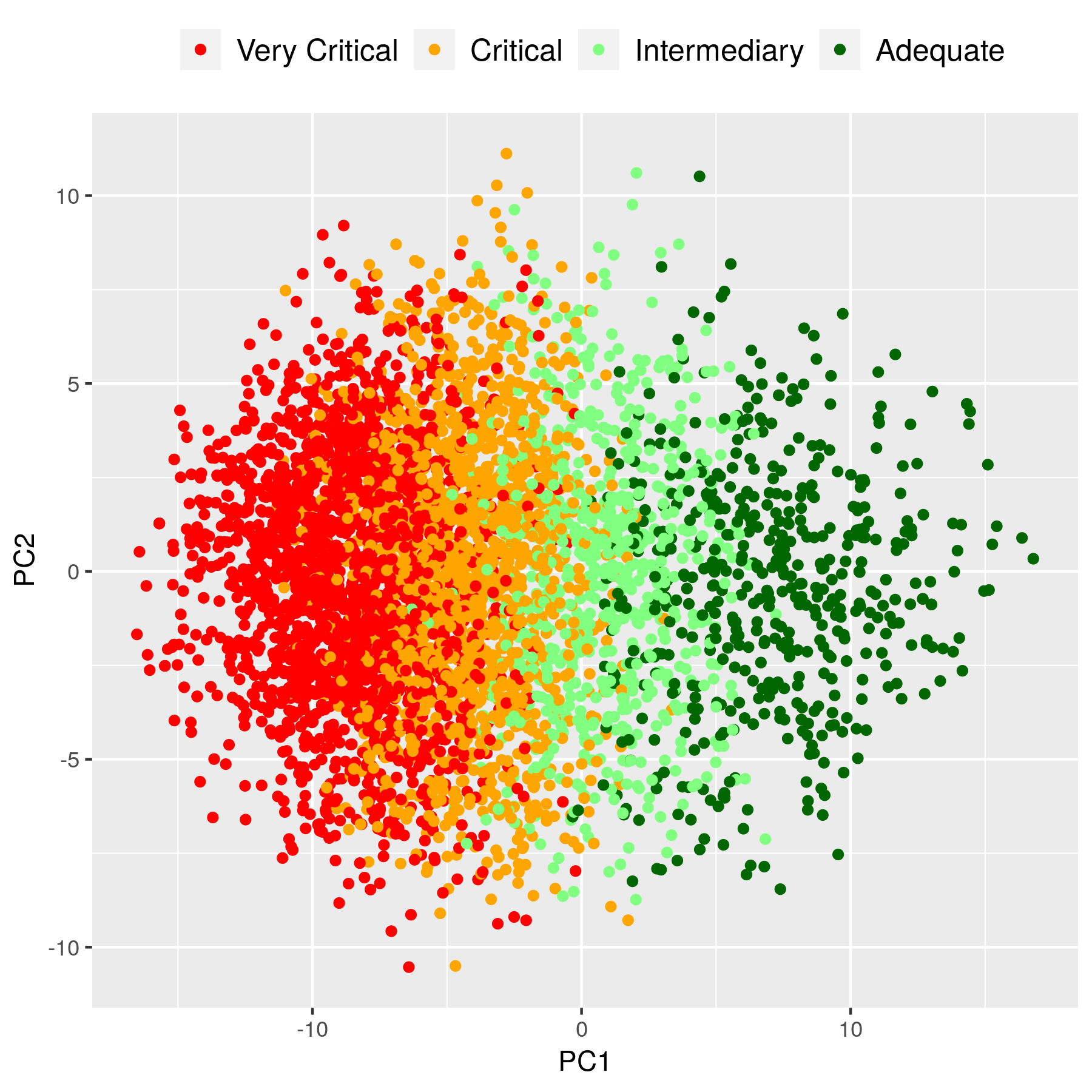}
 \includegraphics[height=7.5cm]{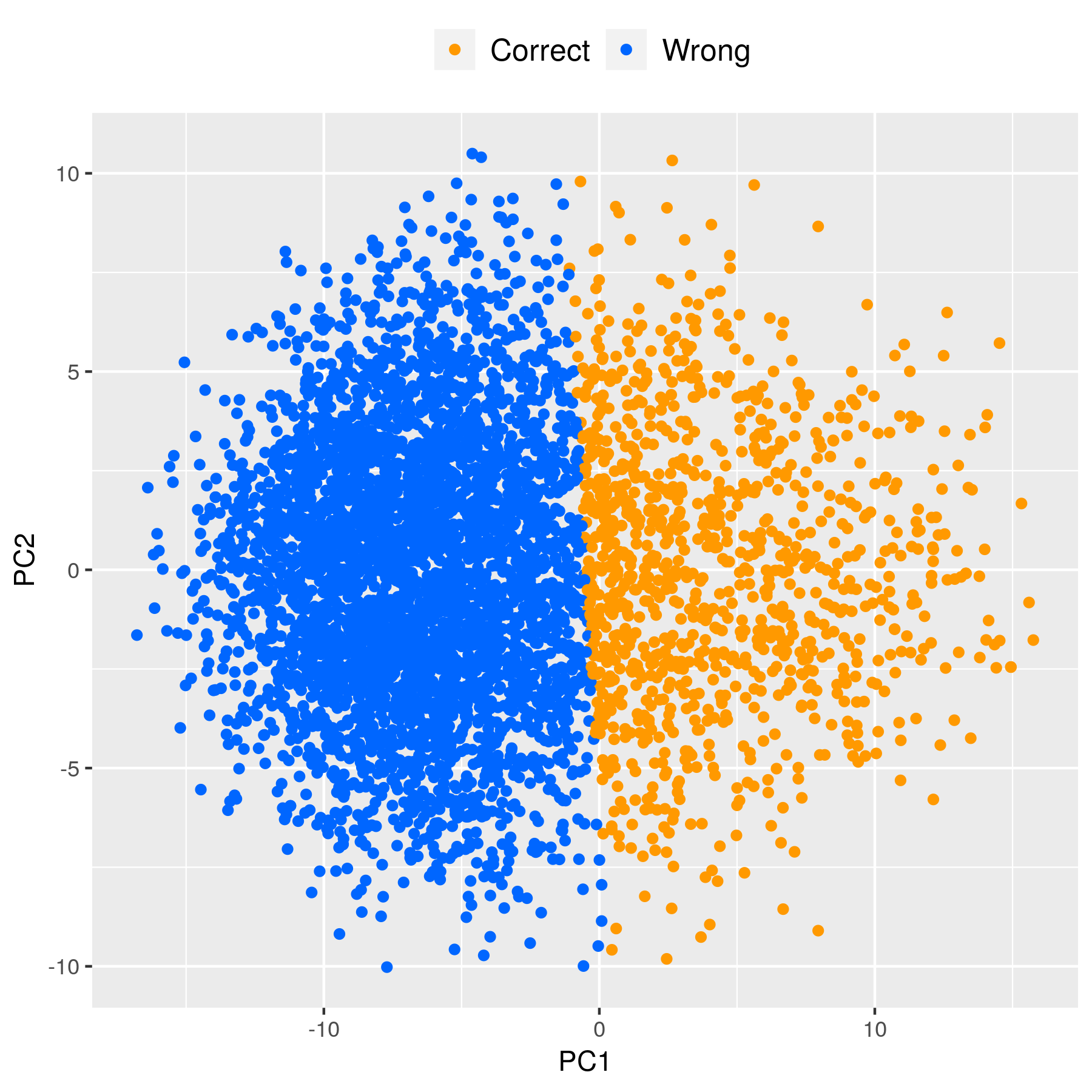}
\input{plots/loadings1.tex}
    \caption{Logistic PCA results of 5K individuals:  two-dimensional scatter plot colored by proficiency (top left) and colored by descriptor $D_{16}$ (top right); relative loadings of the first principal component (bottom). }
    \label{fig:irtSummary}
\end{figure}

Let us consider the SPAECE assessing $\sim 100K$ students at the ending of high school (L3). The test is composed by a set of 24 descriptors from the RM. 
When we apply  LPCA to generate a two-dimensional map, we obtain a picture that carries 
similarities with the IRT scores of SPAECE. In Figure~\ref{fig:irtSummary}, we summarize the observations. The scatter plot at top left exhibits the correlation between 
the IRT scores and PC1. PC2 might correspond to a possible unknown latent trait. The bar plot at the bottom presents the loadings (coordinates)  of PC1 by 
descriptor. They are co-related with the discrimination property of an item. At the top right, we show the profile of correct/wrong answers corresponding 
to descriptor with highest loading in PC1, that is, $D_{16}$. Throughout the paper we clarify  these findings and discuss how to interpret them.

\subsection{Logistic PCA and  UIRT/MIRT}
\label{subsec:LPCA_IRT}
 
\begin{figure}[hptb]
\label{fig:icc}
\centering
\input{plots/logistic_fn.tex}
\caption{Item Characteristic Curve (ICC).}
\end{figure}
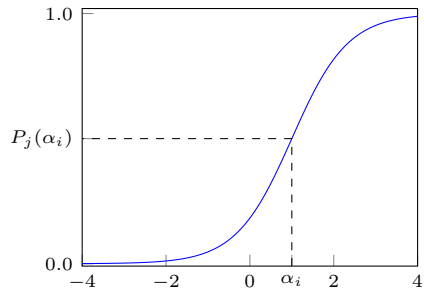

In \emph{Unidimensional Item Response Theory} (UIRT), the Item Characteristic Curve (ICC) 
(see Figure~\ref{fig:icc})
represents the probability that an individual $i$, with a single ability level $\alpha_i$, 
can solve correctly the item~$j$. 

The ICC is a two-parameter logistic model (2PL) given by

\begin{equation}
\displaystyle P(\alpha_i) = \frac{\exp{[\delta_j(\alpha_i - \beta_j)]}}{1+\exp{[\delta_j(\alpha_i - \beta_j)}]},
\end{equation}

\noindent where the parameters $\delta_j$ and $\beta_j$ represent  the \emph{discrimination} 
and \emph{difficulty} of the item, respectively. Notice that if we substitute $-\delta_j\beta_j$ 
with $d_j$, the linear expression turns into the \emph{slope-intercept} form, $\delta_j\alpha_i  + d_j$.
A more general model is the multidimensional two-parameter logistic model
(M2PL)~\cite{Reckase2009} which is given 
by the probability function

\begin{equation}
\displaystyle P_j(\boldsymbol{\alpha_i}) =
    \frac{\exp{(\boldsymbol{\delta_j^T}\boldsymbol{\alpha_i} +
    d_j)}}{1+\exp{(\boldsymbol{\delta_j^T}\boldsymbol{\alpha_i} + d_j)}}.
\end{equation}

\noindent This function takes multiple abilities in a $k\times 1$ vector
$\bs{\alpha_i}$ and mimics the slope-intercept form with the expression
$\bs{\delta_j^T}\boldsymbol{\alpha_i}+d_j$.   Here, $\boldsymbol{\delta_j}$ is a $k \times 1$ vector that 
represents the  \textit{relative discrimination} parameters  of the item~$j$. The scalar $d_j$ parameter is 
not the difficult parameter in the usual sense of a 2PL model because it does not give a unique indicator  
of the difficulty of the item.  Instead, the quotient~$-d_j/\delta_{jl}$ gives the \emph{relative difficulty} 
of the item related to the ability axis~$l$.

By setting $p_{ij}=P_j(\boldsymbol{\alpha_i})$, from
Equation~(\ref{eq:link_function}), we have that:

\begin{align}
\label{eq:connection1}
\theta_{ij}  & = \textrm{logit}\; p_{ij} = \textrm{logit}\;P_j(\boldsymbol{\alpha_i})  = 
    \log
    \left[\frac{P_j(\boldsymbol{\alpha_i})}{1-P_j(\boldsymbol{\alpha_i})}\right] = \nonumber \\[0.2cm]
             & =  \log
             \left[\frac{\dfrac{\exp{(\boldsymbol{\delta_j}^T\boldsymbol{\alpha_i}
             + d_j)}}{1+\exp{(\boldsymbol{\delta_j}^T\boldsymbol{\alpha_i} +
             d_j)}}}{1-\dfrac{\exp{(\boldsymbol{\delta_j}^T\boldsymbol{\alpha_i}
             + d_j)}}{1+\exp{(\boldsymbol{\delta_j}^T\boldsymbol{\alpha_i} +
             d_j)}}}\right] = 
             \log
             \left[\frac{\dfrac{\exp{(\boldsymbol{\delta_j}^T\boldsymbol{\alpha_i}
             + d_j)}}{1+\exp{(\boldsymbol{\delta_j}^T\boldsymbol{\alpha_i} +
             d_j)}}}{\dfrac{1}{1+
             \exp{(\boldsymbol{\delta_j}^T\boldsymbol{\alpha_i} +
             d_j)}]}}\right] \nonumber \\[0.2cm]
             & = \log \left[
                 \frac{\exp{(\boldsymbol{\delta_j}^T\boldsymbol{\alpha_i} +
                 d_j)}}{1+\exp{(\boldsymbol{\delta_j}^T\boldsymbol{\alpha_i}
                 + d_j)}}\cdot
                 [1+\exp{(\boldsymbol{\delta_j}^T\boldsymbol{\alpha_i} +
                 d_j)}]\right]\nonumber \\
            & = \log  \{ \exp{(\boldsymbol{\delta_j}^T\boldsymbol{\alpha_i} + d_j)} \} = \nonumber \\
            & = \boldsymbol{\delta_j}^T\boldsymbol{\alpha_i} + d_j
\end{align}

Recall from Equation (\ref{eq:projection}) that LPCA  provides a vector
$\bs{\mu}$ $\in  \R^d$ and an orthogonal matrix $U_{d\times k}$ such that
$\bs{\theta_{i}}$ is close to $\bs{\tilde {\theta}_i}$. Therefore, it is easy to see that 

\begin{equation}
\label{eq:connection2}
\theta_{ij} =  \bs{u_j }\bs{\psi_i} + \bs{\mu_j}
\end{equation}

\noindent where $\bs{\psi_i} = U^T (\bs{\tilde{\theta}_i}^T-\bs{\mu})$ and $\bs{u_j}$ is the $j$th row of  matrix $U$. 

From Equations~(\ref{eq:connection1}) and (\ref{eq:connection2}) we obtain a connection between  UIRT/MIRT and  LPCA. 
The relative discrimination vector~$\bs{\delta_j}$ in Equation~(\ref{eq:connection1}) expresses how item $j$ can differentiate 
among examinees  with different abilities. It corresponds to the principal component $\bs{u_j}$ in Equation~(\ref{eq:connection2}).
The  scalar $d_j$ represents the $j$-th coordinate of the translation
vector~$\bs{\mu}$ and the vector of multiple 
abilities  $\bs{\alpha_i}$ correspond to the resulting vector $\bs{\psi_i}$ in the LPCA model.


\subsection{A Geometric Representation}
\label{sub:geometric_interpretation}

\begin{figure}[b]
    \centering  
    \includegraphics[height=5.5cm]{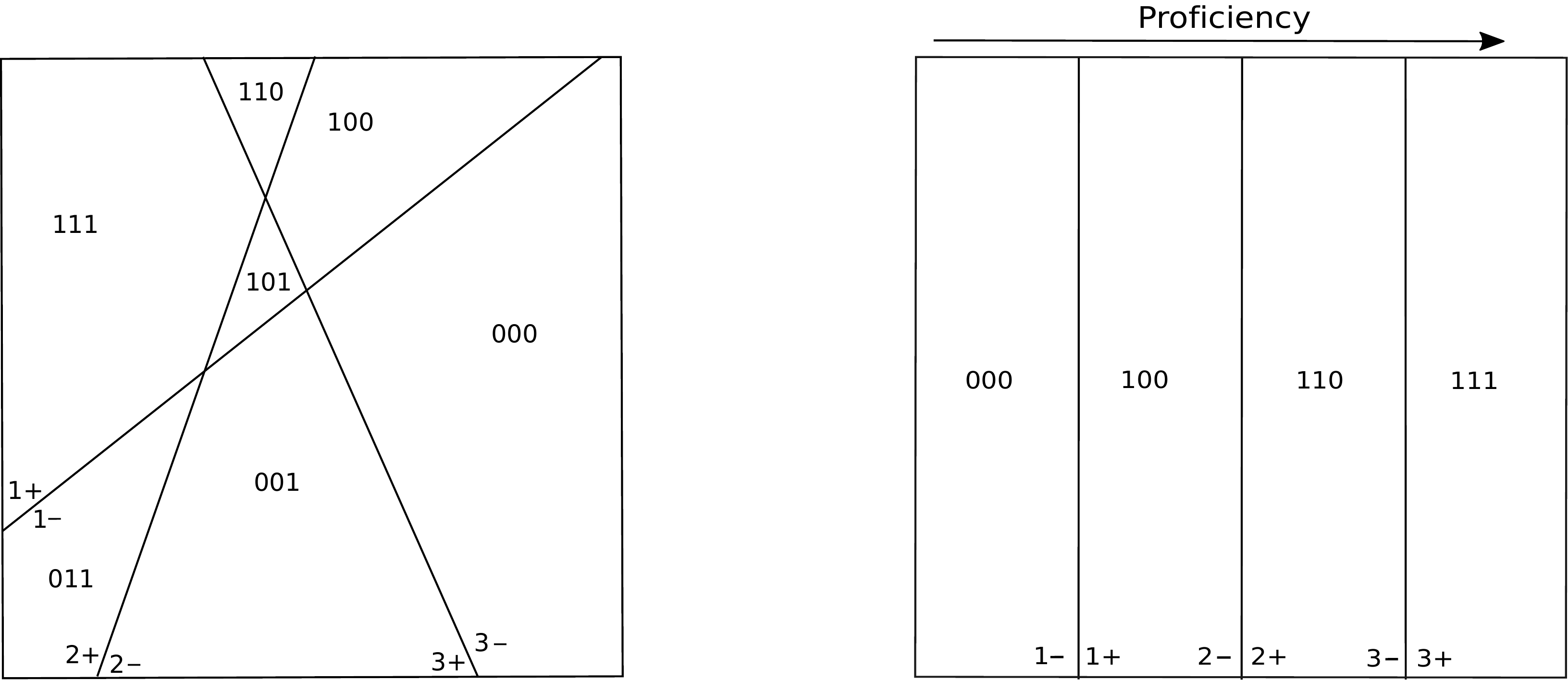}
    \caption{Examples of IPR. Three lines splitting the plane into seven regions (left) and an ideal IRT configuration (right). }
    \label{fig:lines_drawing}
\end{figure}

Another equivalent model to LPCA is the following geometric problem:
represent the rows of the data matrix as points and the columns as
hyperplanes in low-dimensional Euclidean space $\mathbb{R}^k$. Rows $i$
correspond  to points $\bs{a_i}$ and the columns are represented as
hyperplanes $(\bs{b_j}, c_j)$ where $\bs{b_j}$ is a vector of slopes and
$c_j$ is a scalar intercept. As usual, the parameter $k$ represents the dimension.

The solution of the problem consists in construct a drawing in such a way that points $\bs{a_i}$ for which $x_{ij}=1$ (here $X = \{x_{ij}\}$ is a binary data) should be in one side of the plane and the points for which $x_{ij}=0$ should be on the other side. In algebraic terms, we want to find a solution to the system of strict inequalities

\begin{subequations}
\begin{align}
\label{eqn:inequalities0}
\bs{b_j}^T\bs{a_i} > c_j  \textrm{, for } x_{ij}=1\,\,  \\
\label{eqn:inequalities1}
\bs{b_j}^T\bs{a_i} < c_j  \textrm{, for } x_{ij}=0. 
\end{align}
\end{subequations}

\noindent We will call such geometric model as {\it Inner Product Representation} (IPR). In general,  the system of inequalities 
(\ref{eqn:inequalities0}) and (\ref{eqn:inequalities1}) will not have an
exact solution. Therefore, it is necessary to find an approximate solution in the sense that it minimizes the scaled deviance as defined in Equation~(\ref{eq:scaledDeviation}). We refer the reader to \cite{deLeeuw2006} for more details on the approximating solution. 

Figure \ref{fig:lines_drawing} (left) shows an example of solution with
three lines that split the plane into seven regions. Figure
\ref{fig:lines_drawing} (right) shows a theoretical  configuration for IRT:
every individual will get $1$ for items with difficulty below than  the
proficiency of the individual and $0$ otherwise. Notice that the solution
for this configuration has vertical parallel lines and the proficiency of
the individuals increases as the regions goes from left to right. From this
example, we can infer that  the vector of slopes $\bs{b_j}$  express the discrimination of the item. 
More precisely, let us say that there is a solution satisfying inequalities (\ref{eqn:inequalities0})
and (\ref{eqn:inequalities1}). Then, by setting $\bs{\delta_j}:= \bs{b_j}$, $\bs{\alpha_i} = \bs{a_i}$ and $d_j:= -c_j$, 
we have that:

\begin{subequations}
\begin{align}
\label{eqn:inequalities2}
    P_j(\bs{\alpha_i}) > 0.5  \,\,\, \textrm{ if }  \,\,\,
    \bs{\delta_j}^T\bs{\alpha_i} > -d_j \\
\label{eqn:inequalities3}
    P_j(\bs{\alpha_i}) = 0.5  \,\,\,  \textrm{ if } \,\,\,
    \bs{\delta_j}^T\bs{\alpha_i} = -d_j \\
\label{eqn:inequalities4}
    P_j(\bs{\alpha_i}) < 0.5  \,\,\, \textrm{ if } \,\,\,
    \bs{\delta_j}^T\bs{\alpha_i} < -d_j 
\end{align}
\end{subequations}

\noindent One can easily check that the level sets  of $P_j$ are hyperplanes. Equation~(\ref{eqn:inequalities3}) 
represents the hyperplane $(\bs{\delta_j}, -d_j)$ and it corresponds exactly to the level set $0.5$.

\subsection{Equivalence Table}

So far, in this section, we synthesized the theory of three different models (LPCA, UIRT/MIRT and IPR) for 
representing multidimensional data in a lower dimensional space. More importantly, we have established a connection among them. 
In Table~\ref{tb:equivalence}, we summarize the equivalence of the parameters through the models which provides different insights
that will help us to explore and understand data in the applications.

\begin{table}[hptb]
\centering
\begin{tabular}{p{2.5cm}|p{5.5cm}|p{4.5cm}}

LPCA     &   UIRT/MIRT & IPR \\
\hline
vector $\bs{u_j}$ &   vector of relative discriminations  $\bs{\delta_j}$ & vector of slopes $\bs{b_j}$\\
\hline
coordinate $\mu_j$ &  intercept $d_j$    &  intercept $c_j$ \\
\hline
vector $\bs{\psi_i}$ & vector of multiple abilities $\bs{\alpha_i}$ & point position $\bs{a_i}$\\
\end{tabular}
\caption{\label{tb:equivalence}Equivalence table of parameters through the models LPCA, UIRT/MIRT and IPR.}
\end{table}

%% file: plots/loadings1.tex
\pgfplotsset{
tick label style={font=\scriptsize}, 
compat=newest, 
width=17cm, height=3.5cm, 
label style={font=\scriptsize}, 
}
\begin{tikzpicture}
      \begin{axis}[ybar, 
      enlarge x limits=0.05,
      enlarge y limits=0.10,
      xtick = data,
      symbolic x coords= {$D_{16}$, $D_{76}$, $D_{42}$, $D_{57}$, $D_{52}$, $D_{78}$, $D_{40}$, $D_{19}$, $D_{65}$, $D_{55}$, $D_{28}$, $D_{58}$, $D_{54}$, $D_{72}$, $D_{50}$, $D_{64}$, $D_{53}$, $D_{51}$, $D_{67}$, $D_{24}$, $D_{20}$, $D_{56}$, $D_{49}$, $D_{71}$}, 
      ylabel=Loadings in \%,
      xlabel = Descriptors,
      ] 
      \addplot [draw=blue,fill=blue, opacity=0.5] 
      coordinates {($D_{16}$, 13.01) ($D_{76}$, 11.08) ($D_{42}$, 9.41) ($D_{57}$, 8.42) ($D_{52}$, 8.31) ($D_{78}$, 6.80) ($D_{40}$, 5.07) ($D_{19}$, 3.91) ($D_{65}$, 3.73) ($D_{55}$, 3.54) ($D_{28}$, 3.07) ($D_{58}$, 3.01) ($D_{54}$, 2.87) ($D_{72}$, 2.83) ($D_{50}$, 2.62) ($D_{64}$, 2.24) ($D_{53}$, 2.03) ($D_{51}$, 1.76) ($D_{67}$, 1.58) ($D_{24}$, 1.46) ($D_{20}$, 1.26) ($D_{56}$, 1.03) ($D_{49}$, 0.68) ($D_{71}$, 0.27)};
      
      \addplot[ dashed, black, sharp plot, update limits=true] 
      coordinates {($D_{16}$, 4.16) ($D_{71}$, 4.16)} 
      node[above] at (axis cs:$D_{49}$,4.16) {\scriptsize Mean};
      \end{axis} 
 \end{tikzpicture}

%% file: plots/logistic_fn.tex
\pgfplotsset{
tick label style={font=\scriptsize}, 
compat=newest, 
width=6cm, height=5cm, 
label style={font=\scriptsize}, 
}
\begin{tikzpicture}[
declare function={myfun(\a,\b,\c)=1/(1+exp(-\a*(\c-\b)));},
]
    \begin{axis}%
    [
       enlarge y limits=0.01,
        xmin=-4,
        xmax=4,
        xtick pos=left,
        ytick pos=left,
        xtick={-4,-2, 0, 1, 2, 4},
        ytick={0.0,0.5,1.0},
        ymax=1.01,
        yticklabel style={
        /pgf/number format/precision=1,
        /pgf/number format/fixed,
        /pgf/number format/fixed zerofill},
        yticklabels={$0.0$,$P_j(\alpha_i)$,$1.0$},
        xticklabels={$-4$,$-2$,$0$,$\alpha_i$,$2$,$4$},
    ]
       \addplot%
        [
            blue,%
            mark=none,
            samples=100,
            domain=-4:4,
        ]
        (x, {myfun(1.5, 1.0, x)})
        coordinate [pos=5/8] (A);
        \draw[dashed](1, -0.2) -- (1, {myfun(1.5, 1.0, 1.0)}) -- (-4,  {myfun(1.5, 1.0, 1.0)});
      \end{axis} 
\end{tikzpicture}

%% file: applications.tex

\section{Applications}
\label{sec:applications}

Our applications are focused on representing data in two dimensions. We will
show applications on visualizing colored biplots which highlight properties that users wish to see from the population. We will also perform visual analysis on SPAECE which shows how these tools can be used to perform data exploration.

\subsection{Pre-processing Data Set }

Since 2008, with a three-parameters model, SPAECE  has been adopting an IRT
based methodology for tests elaboration and data analysis. In the end of the
school year every student at levels L1, L2 and L3 are submitted to
Mathematics and Portuguese assessment. Our analysis will focus on  the Mathematics assesments applied from 2016 to 2018 at the level L3.

 The test that is given to each examinee  contains two distinct blocks with
 13 items, totalizing 26 items. These blocks are assembled through the {\it
 balanced incomplete block design} (BIBD)~\cite{Montgomery2003}. 
In each test, there are at most two items assigned to the same descriptor\footnote{Due to confidentiality reasons, we were allowed to retrieve only the corresponding descriptor of the item.}. We feed the input of the LPCA algorithm with a table where the rows correspond to the examinee-test and the columns correspond to the math descriptors (see appendix ~\ref{ap:referenceMatrix}). Depending on the number of items per descriptor in the test, the columns will be filled by $NA$ (not available) or the probability of getting a correct answer. So, if the examinee gets one/two correct answer out of one/two items the probability is $1$; one correct answer out of two items results in $1/2$; no correct answers results in $0$. In Table~\ref{tb:input} we show the first samples filled of a typical input.

In summary, the variables under analysis are the descriptors as listed in the RM. For each examinee-test, the value of each descriptor corresponds to the rate of 
correct items assigned to it.   

We emphasize that, although the previous section has assumed a binary matrix as input, one can easily check that the theory is still valid for values from 0 to 1.  Moreover, LPCA can also manage missing data, that is, descriptors with values assigned  as not available ($NA$).

\newcolumntype{g}{>{\columncolor{Gray}}c}
\begin{table}[t]
\footnotesize
\centering
\begin{tabular}{|g|c|c|c|c|c|c|p{2cm}|c|c|}
\hline
\rowcolor{Gray}
examinee & $D_{16}$ & $D_{19}$ & $D_{20}$ & $D_{24}$ & $D_{28}$ & $D_{40}$ & \dots & $D_{76}$ & $D_{78}$ \\
\hline
$S_1$ & $0$ & $NA$ & $1/2$ & $NA$ & $NA$ &  $0$ &  \dots & $0$ & $1/2$ \\
\hline
$S_2$ & $0$ & $NA$ & $0$   & $1$  & $0$ &  $0$ &  \dots & $1$ & $1$ \\
\hline
$S_3$ & $0$ & $NA$ & $0$   & $NA$ & $NA$ &  $0$ &  \dots & $1$ & $0$ \\
\hline
$S_4$ & $1/2$ & $NA$ & $0$ & $0$ & $0$  &  $0$ &   & $1$ & $0$ \\
$\vdots$ & $\vdots$  & $\vdots$   & $\vdots$    & $\vdots$   & $\vdots$    & $\vdots$ &  $\ddots$ & $\vdots$  &  $\vdots$ \\
      
\end{tabular}
\caption{\label{tb:input} First rows of the LPCA  algorithm input.}
\end{table}


\subsection{Visualizing LPCA results}

As we run the LPCA algorithm with two principal components we get
the bi-dimensional scatter plot. 
Below, we give examples of color maps that may be useful for analyses. For a better visualization of the figures we reduced the number of individuals from $\sim 100K$  to $7K$ individuals randomly selected. We classify the maps in to groups: {\it ability maps} and {\it social maps}.

\subsection*{Ability Maps}

Ability  maps are primarily concerned with showing the efficiency behavior of the entire population. We  exemplify here two types of maps.

\noindent \textbf{Proficiency Map}: SPAECE provides the IRT proficiency
scores of the individuals in a scale that ranges from $0$ to $500$. This
interval is subdivided into four performance standards: $[0-250]$ is {\it
very critical}; $[251-300]$ is {\it critical}; $[301-350]$ is {\it
intermediate} and $[351-500]$ is {\it adequate}. In
Figure~\ref{fig:irtSummary} (top lef) we show the color map of the
proficiency scores of the assessment in 2018 subdivided in these groups. As
predicted in Section~\ref{subsec:LPCA_IRT}, we can see a strong correlation
between the proficiency scores and PC1 (i.e. the unidimensionality property
of the IRT model), which is confirmed by a Pearson's correlation coefficient
value of $0.92$. It is interesting to see that the  graphic also shows the
density decreasing as we go over the PC1 from left to right. Since the first
SPAECE assessment was applied, the higher concentration of individuals
persists on the left side, i.e., most of the population are located among the {\it very critical} and {\it critical} performance standards. This indicates a poor quality of the public education system delivered by state of Cear\'a. 

\noindent \textbf{Descriptor Map}: The descriptor map is a binary color map that brings insights of the population regarding their performance  per descriptor. For each descriptor $j$ we split the population into two groups: those with probability higher than $0.5$ (i.e. $p_{ij}> 0.5$) of getting descriptor $j$ correctly and those with probability lower than $0.5$ (i.e. $p_{ij} < 0.5$) of getting the descriptor $j$ wrong. In Figure~\ref{fig:maps} (left) we show descriptor $D_{76}$ in 2016.   In this descriptor, most of examenees get correct answers. This means that the descriptor is composed by items with very low difficulty. In Figure~\ref{fig:maps} (middle) we show descriptor $D_{56}$ of 2017  which has an unexpected  behavior. Although a minority of the examenees get correct answers, the slope of the level set line at $0.5$ is nearly horizontal which means that the group of items  composing  this descriptor have poor discrimination. The low discrimination value indicates the irrelevance  of the descriptor to the trait being measured by the test~\cite{furr13}. Even if examinee ``B" can identify correctly circle's equations and
examinee ``B" does it incorrectly, we might not feel confident concluding that
examinee ``A" has a higher proficiency  than does examinee ``B".  In Figure~\ref{fig:irtSummary} (top right) we show the descriptor map of  $D_{16}$. We will discuss more about its importance in Sectionc~\ref{sub:dataAnalysis}.    

\subsection*{Social Maps}

A social map provides visual distributions of social characteristics of the population (e.g. gender, age, family income etc.). Here present the shift map.

\begin{figure}[t!]
    \centering
    \includegraphics[width=5.1cm]{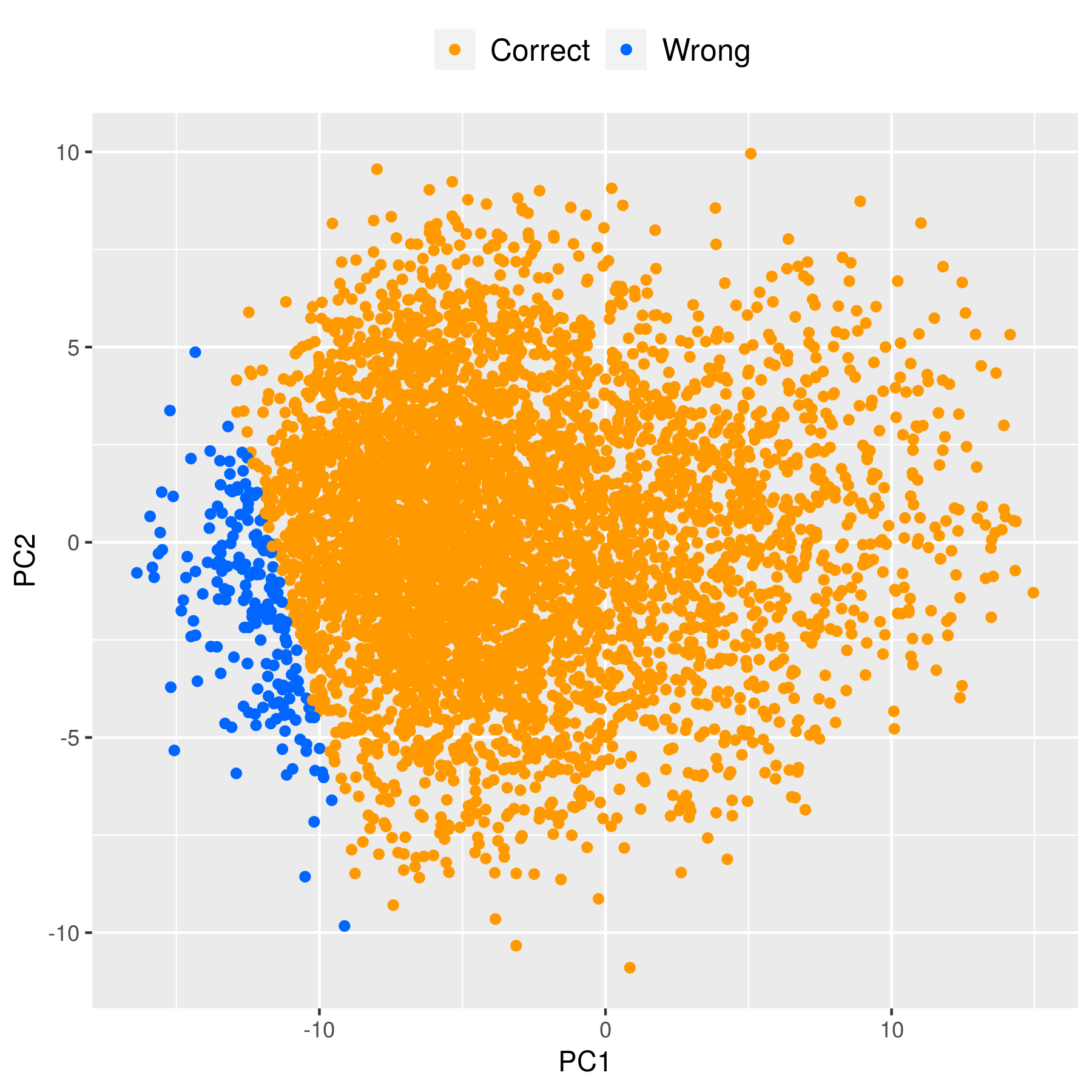}
    \hfill
    \includegraphics[width=5.1cm]{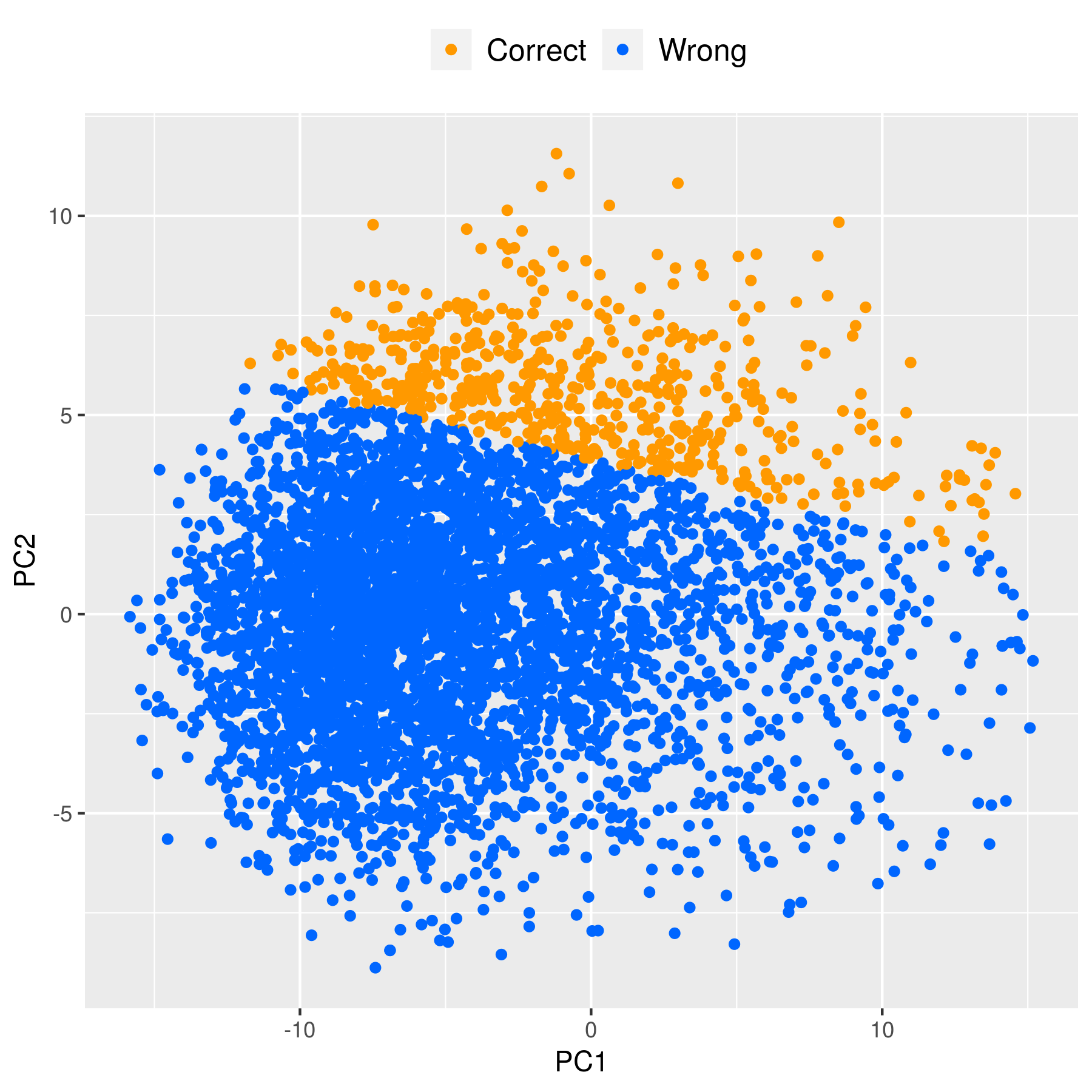}
    \hfill
    \includegraphics[width=5.1cm]{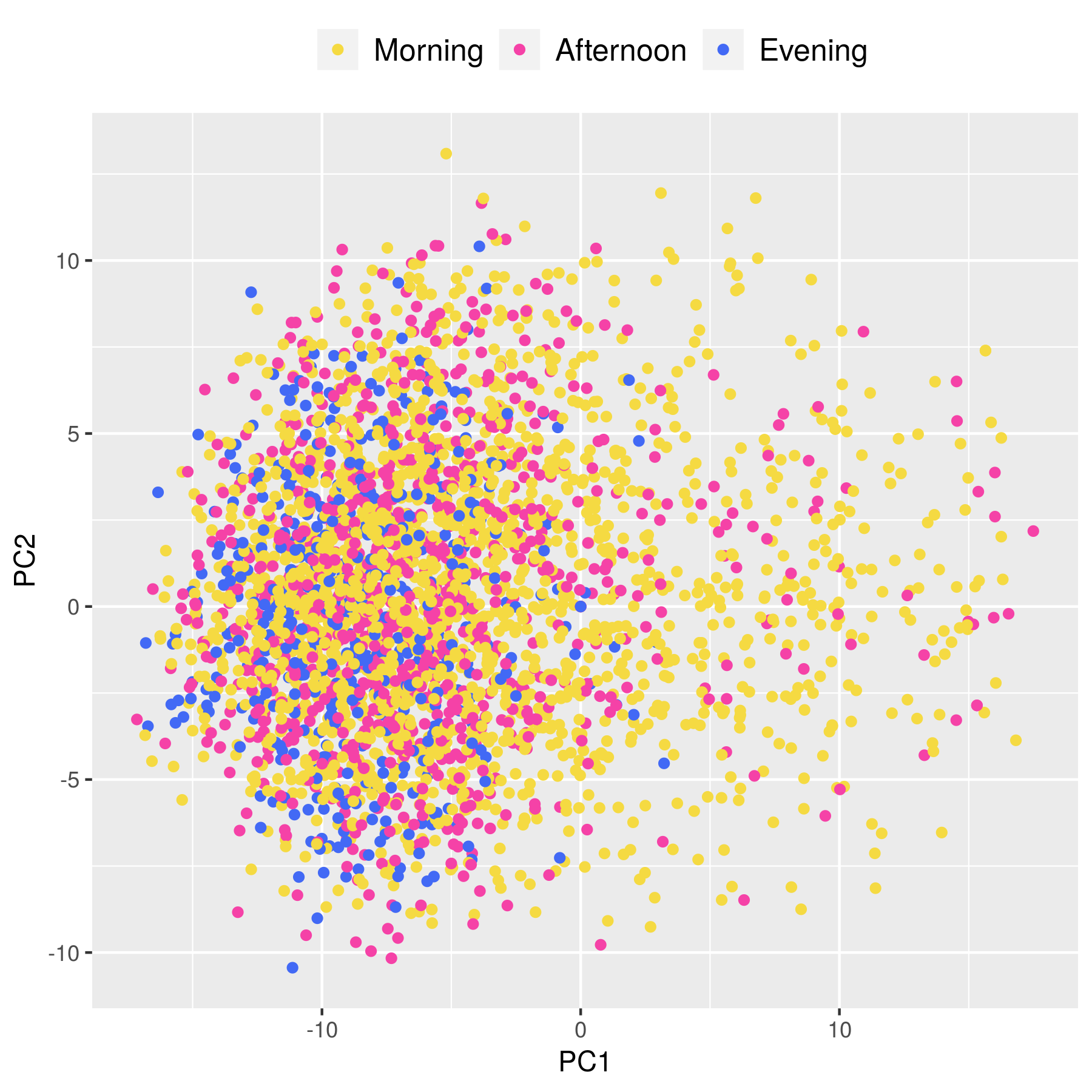}
    \caption{Examples of color maps. Descriptor map of $D_{76}$ and $D_{56}$ (left and middle respectively) and shift map (right).}
    \label{fig:maps}
\end{figure}

\noindent \textbf{Shift Map}: In Figure~\ref{fig:maps}(right) we show the shift map of the population. Observe that the concentration  of the evening group in the left side is higher than the others, that is, this group has a lower proficiency average. There are economic and social reasons involved for this problem. We refer the reader to \cite{unibanco16, ayrton15}  for more details.

\subsection{Example of Visual Analytics}
\label{sub:dataAnalysis}

Despite the poor results of the proficiency in mathematics in state of Cear\'a, there are small islands of excellence among schools in the public education system. Over the period from 2016 to 2018, having a growth of $26.5\%$, the best average proficiency growth was achieved by School~$A$\footnote{Due to confidentiality reasons, we are not allowed to identify the school.}. In this period, the school moved from the {\it very critical} group to the {\it intermediate} performance  group. This growth corresponds to an impressive jump of two levels. Using LPCA and the visualization tools described in this paper, we investigate this outstanding result in more details. 

\begin{figure}[tpb]
\centering
\begin{tabular}{|c|c|c|}
    \hline
     $2016$ & $2017$ & $2018$ \\
    \hline
     &  & \\
    \input{plots/loadings2016.tex} &
    \input{plots/loadings2017.tex}&
    \input{plots/loadings2018.tex}\\
    \includegraphics[width = 4.8cm]{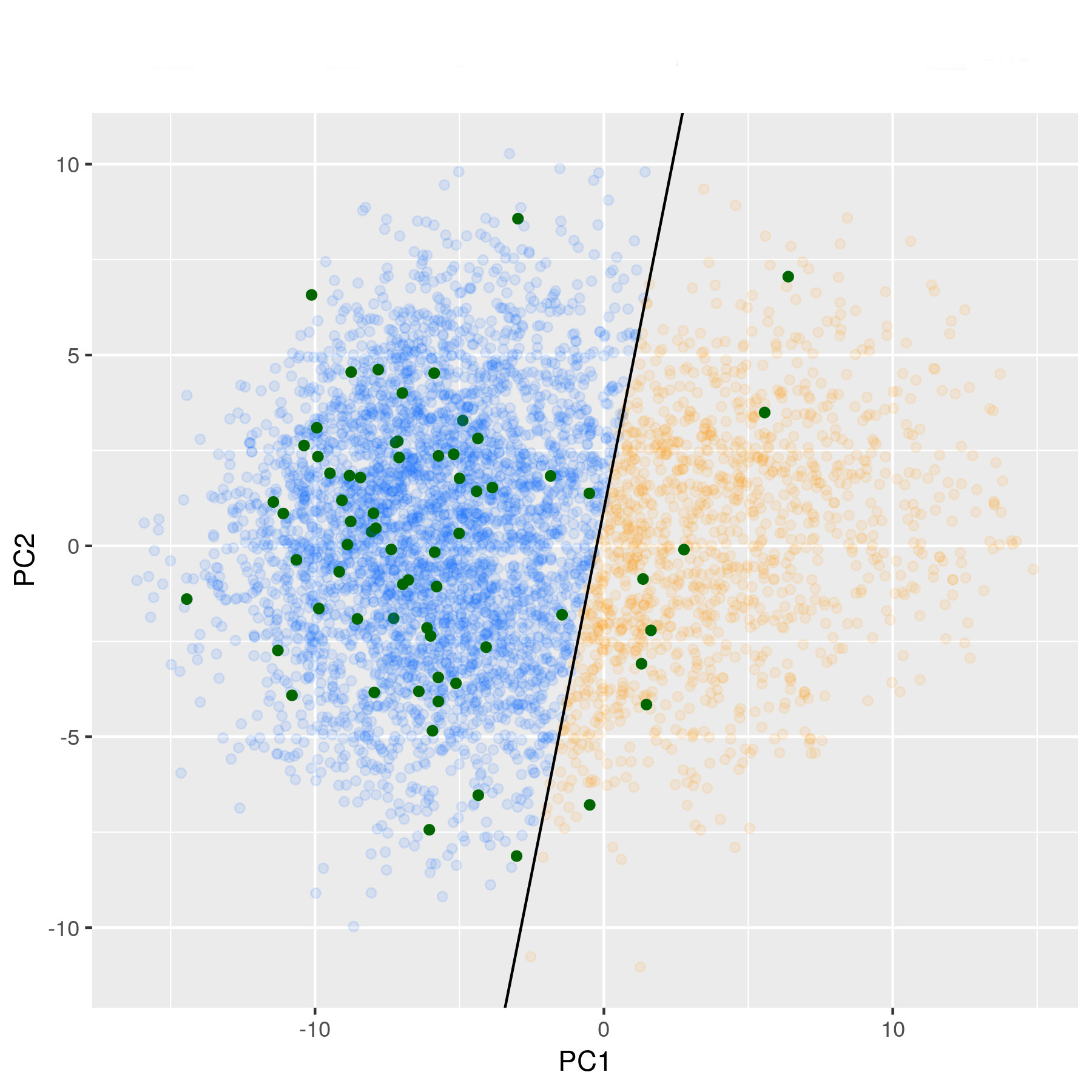} &
    \includegraphics[width = 4.8cm]{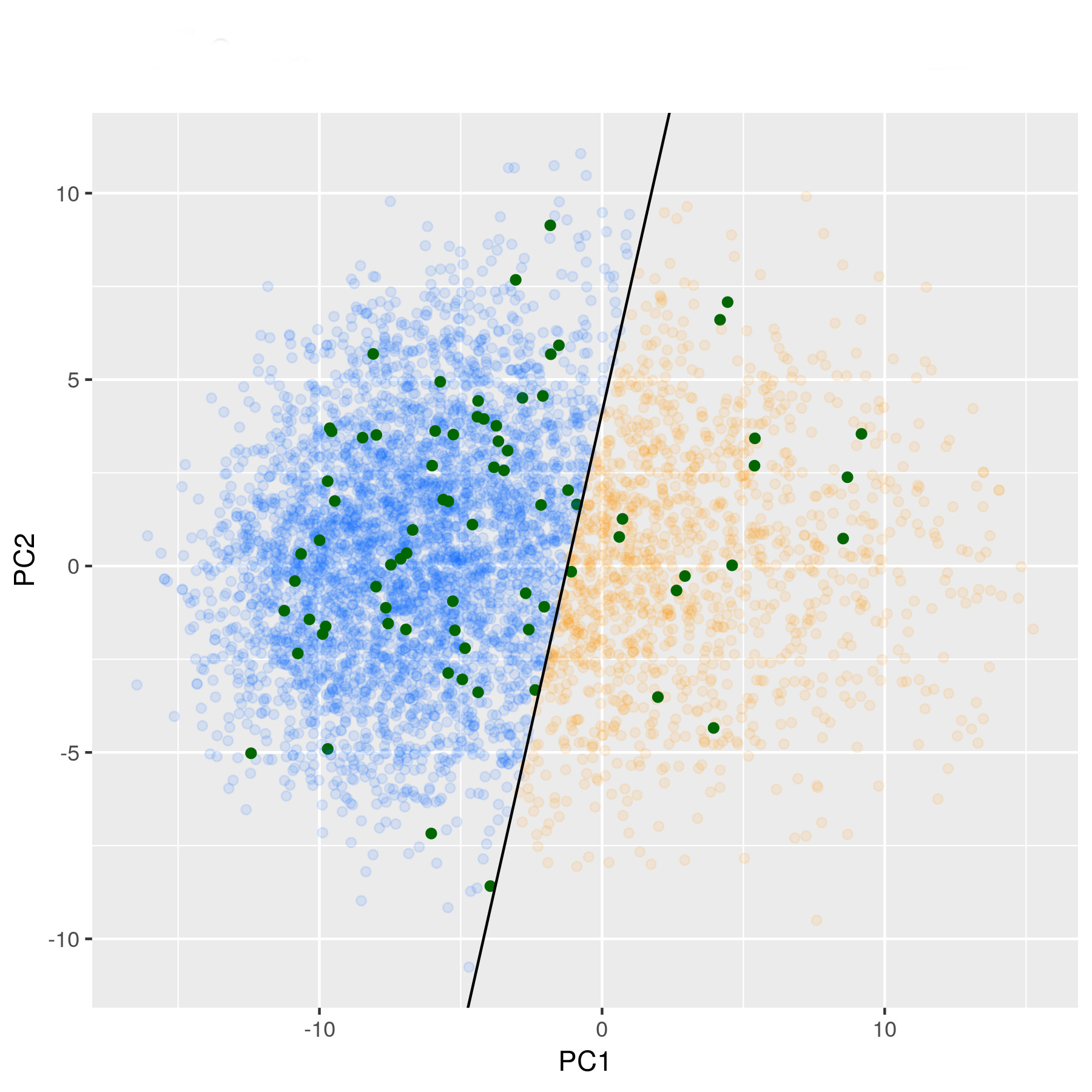} &
    \includegraphics[width = 4.8cm]{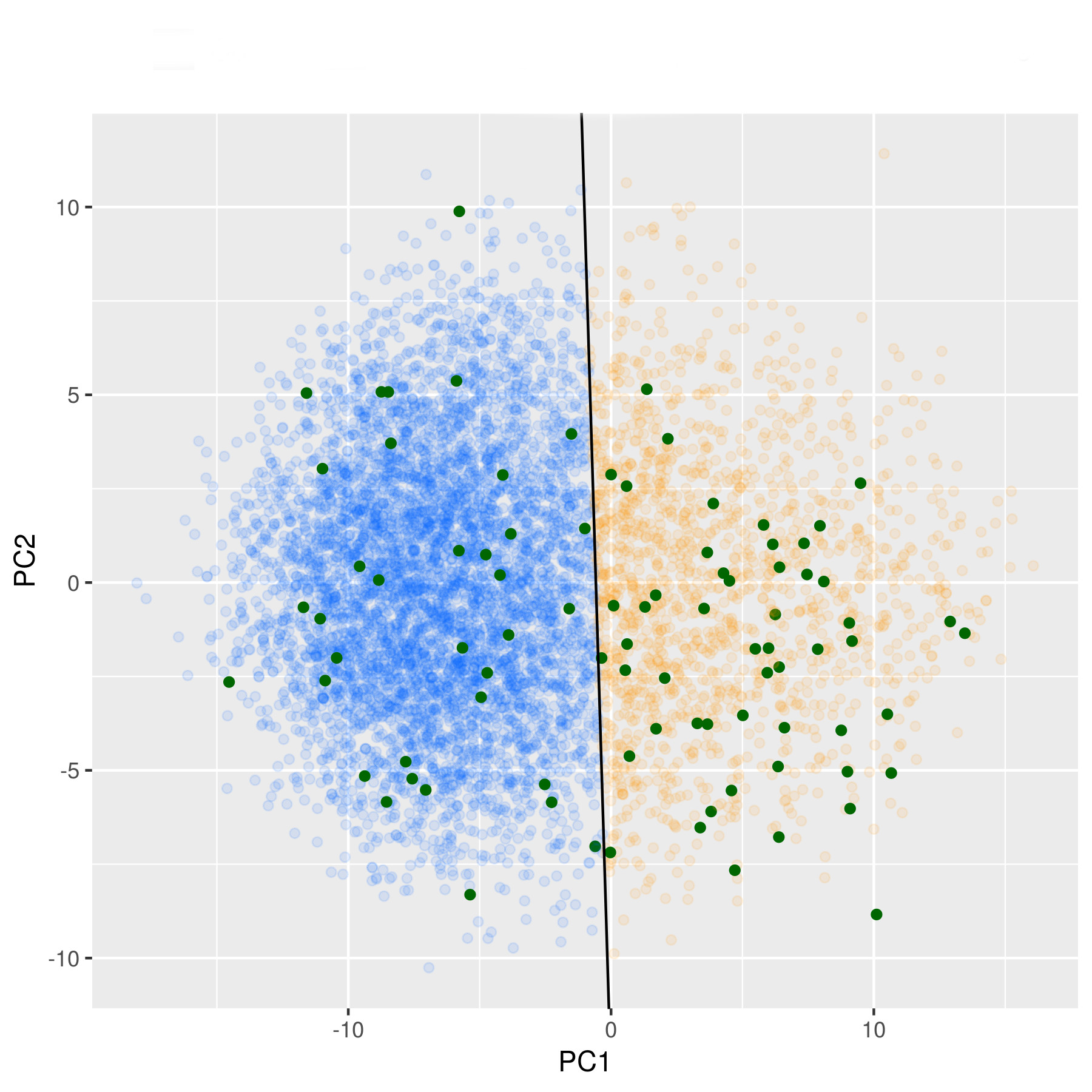}\\
    \includegraphics[width = 4.8cm]{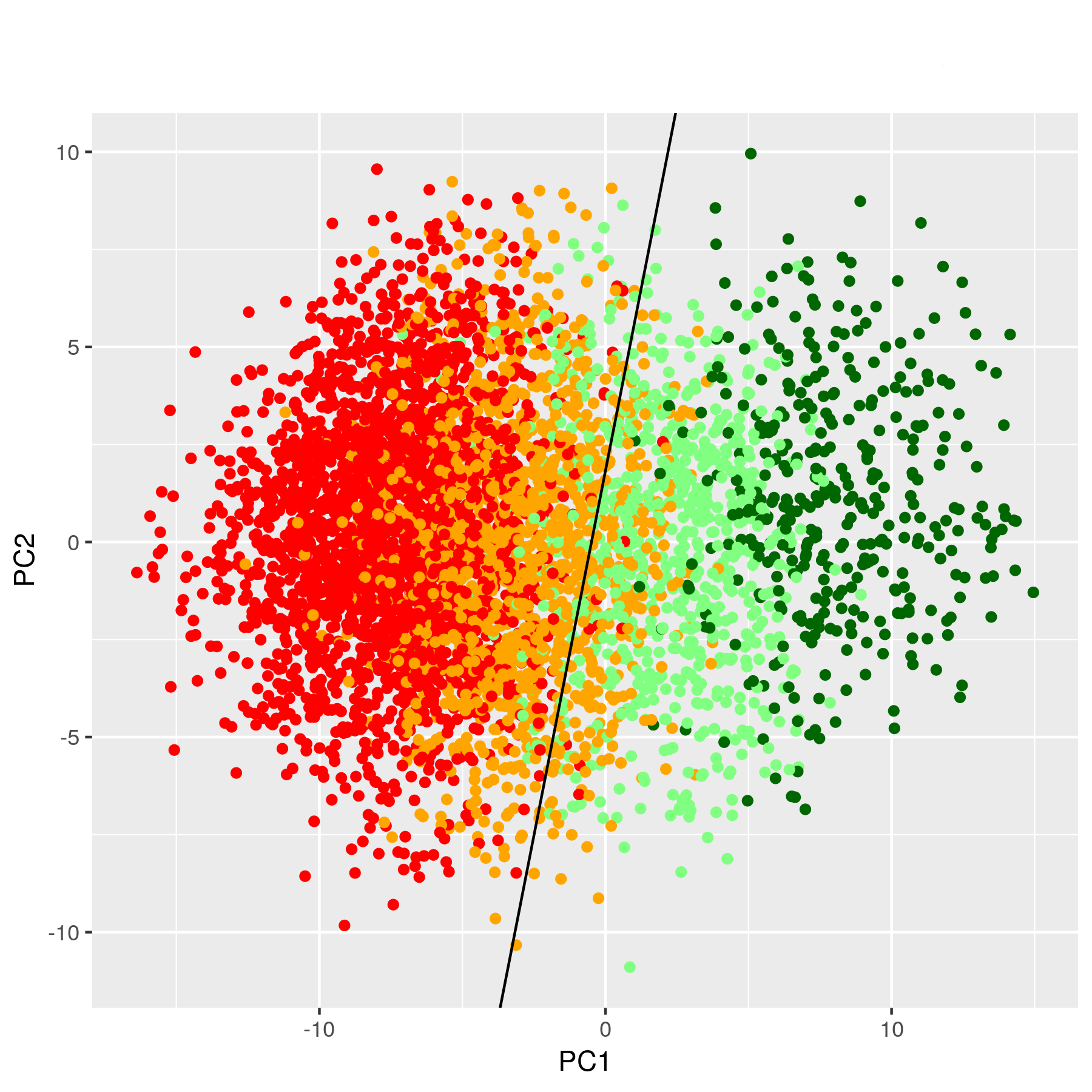} &
    \includegraphics[width = 4.8cm]{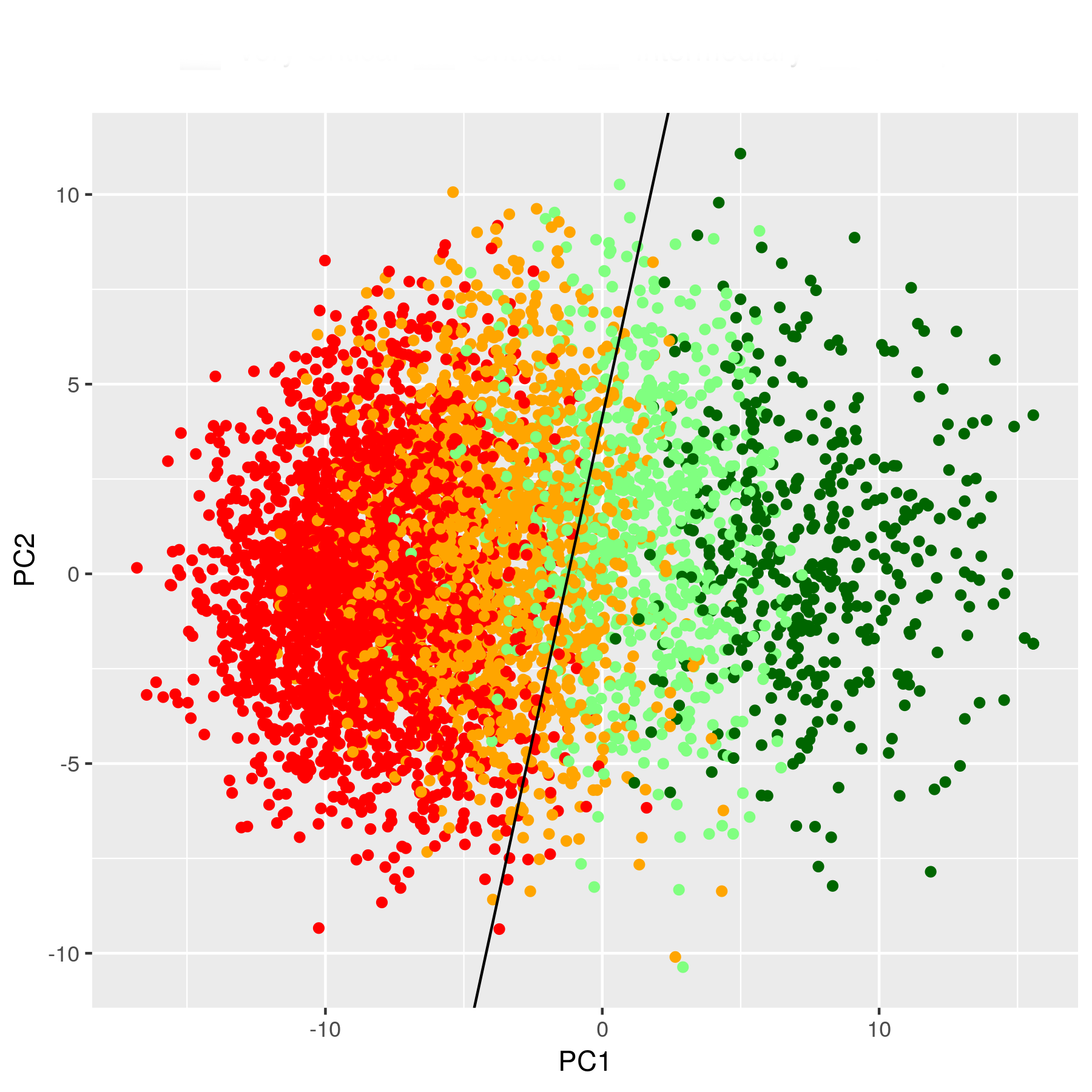} &
    \includegraphics[width = 4.8cm]{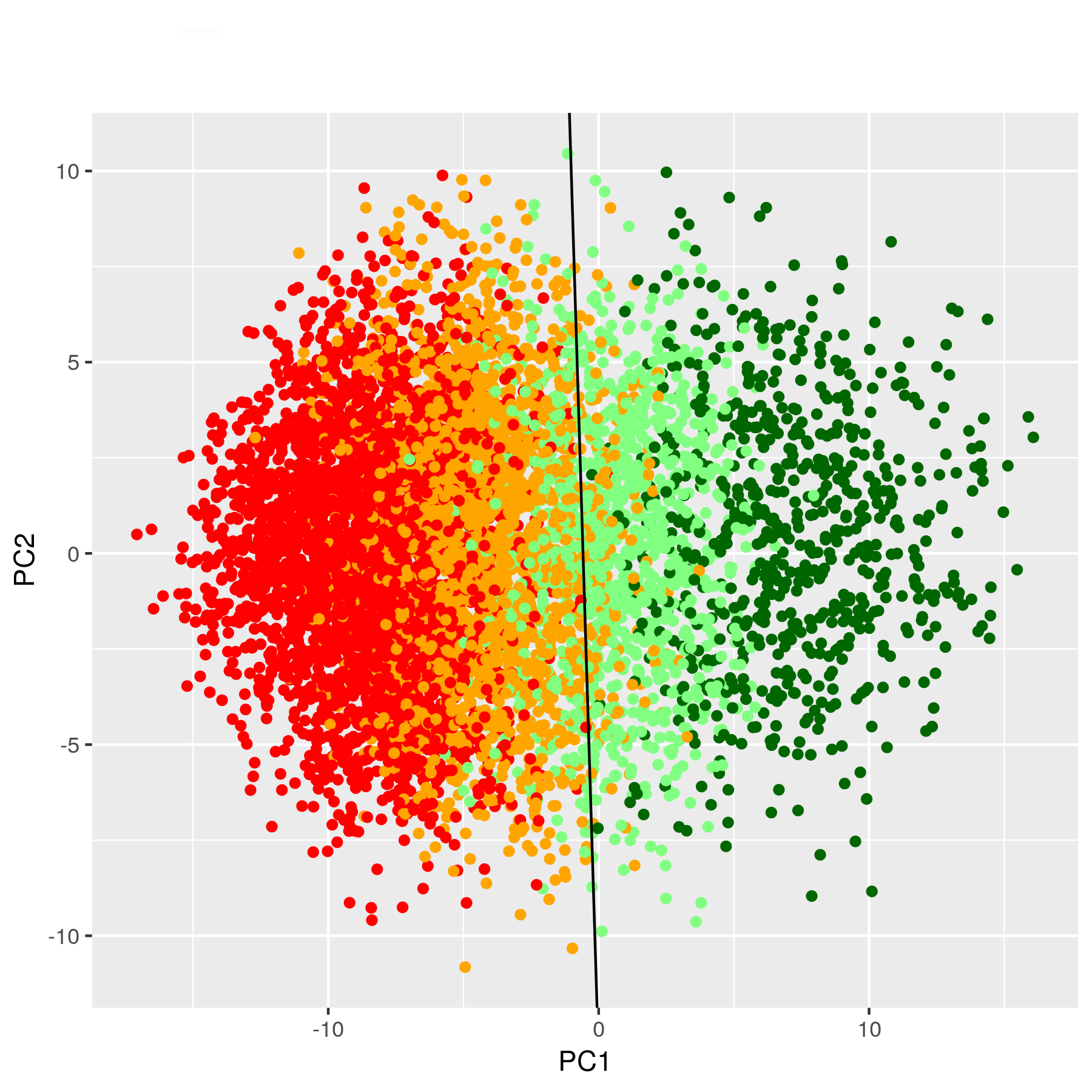}\\
    \hline
\end{tabular}
    \caption{Proficiency growth of School $A$ from 2016 to 2018. First three loadings of PC1 (top row), descriptor map of $D_{16}$ (middle row) and proficiency map (bottom row). }
    \label{fig:dataExploration}
\end{figure}

In Figure~\ref{fig:dataExploration} (top row), we plot the three highest loadings of PC1.  Notice that all plots are headed by descriptor $D_{16}$. Moreover, its values represent at least three times the mean of all 24 descriptors. Recall that, from the equivalence Table~\ref{tb:equivalence}, $D_{16}$ corresponds to the best discrimination's parameter in IRT. 

In  Figure~\ref{fig:dataExploration} (middle row), we show the descriptor
maps of $D_{16}$  where the examinees of School~$A$ are represented by dark
dots. In these maps we observe that all level set lines at $0.5$  are
nearly vertical. Again, as shown in Table~\ref{tb:equivalence}, this
confirms the equivalence between the IRT discrimination's parameter and the
IPR slope's parameter. We observe that there is a strong  migration of dark dots from the ``left side"  to the  ``right side" of the level set $0.5$. If we compare the descriptor maps with the proficiency maps Figure~\ref{fig:dataExploration} (bottom row), we find that the level set lines at $0.5$  are located right between the {\it critical} and {\it intermediate}   groups in the performance standards. 

 More important than visualizing graphs,  is understand why $D_{16}$  has all these peculiar properties as described above.  This descriptor is directly connected with a large research area called {\bf number sense}, a topic in mathematics education which has been developed in the last few decades ~\cite{dahaene11, russell99}. Roughly speaking,  number sense refers to an individual general comprehension
with regard to numbers and flexibility in using the operations for making mathematical
discernment. It is the result of mathematical experience
where students could employ their sense in understanding circumstances involving numbers~\cite{Maghfirah2018}.

\begin{figure}[]
    \centering
    \input{plots/growthSchoolA.tex}
    \caption{Correct answers of School~$A$ per descriptor from 2016 to 2018.}
    \label{fig:numberSenseDependence}
\end{figure}
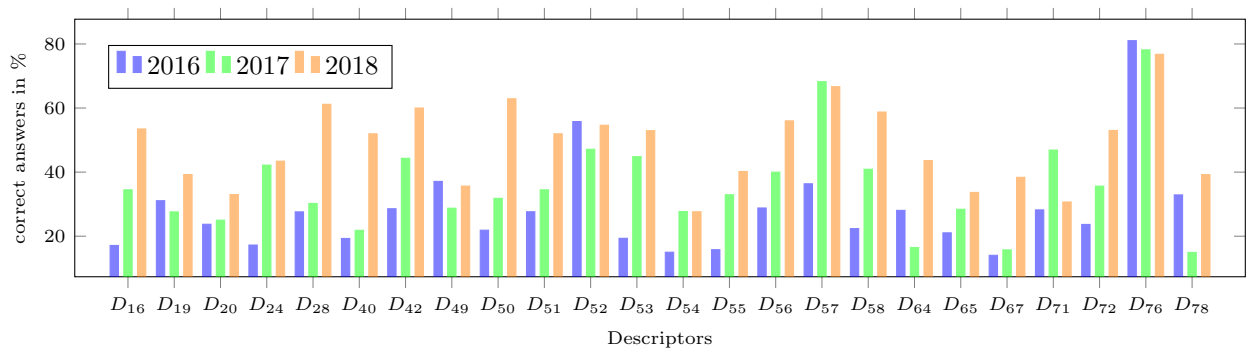

 Number sense is extremely necessary for  individuals to be successful in the other descriptors listed in the Reference Matrix.   For example, to solve a problem corresponding to descriptor $D_{55}$, the examenee needs to understand that the inclination of a straight line is the fraction ${\Delta y}/{\Delta x}$, i.e., the ratio between the the variations of $y$- and $x$- axes in the corresponding linear function. As a matter of fact, a great many people find fractions very difficult to learn because their cortical machinery resists such a counterintuitive concept~\cite{dahaene11} and an examinee tends to fail in this example if he did not understand well how to manipulate ratios and proportions. In Figure~\ref{fig:numberSenseDependence}, we show the plots of the number of correct answers by School~$A$ where  most of the other descriptors follows the growth~trend~of~$D_{16}$.
 
 In interviewing the  School $A$'s math teachers, we checked that they made an special effort in recovering their students with disabilities in number sense. The discussion of the  teaching methodology employed by them is out of the scope of this paper, however, the idea of focusing in number sense could be successfully reproduced by schools with similar conditions in the public education system of Cear\'a. By similar conditions, we mean schools with {\it very critical} performance. As  a result, the students may surpass the ``barrier" line of descriptor $D_{16}$ and  migrate from an average performance in the {\it very critical} group  to the {\it intermediate} group.          
  
%% file: plots/loadings2016.tex
\pgfplotsset{
tick label style={font=\scriptsize}, 
compat=newest, 
width=5cm, height=3.5cm, 
label style={font=\scriptsize}, 
}
\begin{tikzpicture}
      \begin{axis}[ybar, 
      enlarge x limits=0.3,
      ymin=0,
      ymax=17,
      xtick = data,
      symbolic x coords= {$D_{16}$, $D_{57}$, $D_{42}$}, 
      ylabel=Loadings in \%,
      bar width = 0.5cm,
      ] 
      \addplot [draw=blue,fill=blue, opacity=0.5] 
      coordinates {($D_{16}$, 15.87) ($D_{57}$, 13.98) ($D_{42}$, 10.76)} ;
      
      \addplot[ dashed, black, sharp plot, update limits=true] 
      coordinates {($D_{16}$, 4.16) ($D_{42}$, 4.16)}; 
      \end{axis} 
 \end{tikzpicture}

%% file: plots/loadings2017.tex
\pgfplotsset{
tick label style={font=\scriptsize}, 
compat=newest, 
width=5cm, height=3.5cm, 
label style={font=\scriptsize}, 
}
\begin{tikzpicture}
      \begin{axis}[ybar, 
      enlarge x limits=0.3,
      ymin=0,
      ymax=17,
      xtick = data,
      symbolic x coords= {$D_{16}$, $D_{42}$, $D_{52}$}, 
      ylabel=Loadings in \%,
      bar width = 0.5cm,
      ] 
      \addplot [draw=blue,fill=blue, opacity=0.5] 
      coordinates {($D_{16}$, 14.17) ($D_{42}$, 11.61) ($D_{52}$, 9.22)} ;
      
      \addplot[ dashed, black, sharp plot, update limits=true] 
      coordinates {($D_{16}$, 4.16) ($D_{52}$, 4.16)}; 
      \end{axis} 
 \end{tikzpicture}

%% file: plots/loadings2018.tex
\pgfplotsset{
tick label style={font=\scriptsize}, 
compat=newest, 
width=5cm, height=3.5cm, 
label style={font=\scriptsize}, 
}
\begin{tikzpicture}
      \begin{axis}[ybar, 
      enlarge x limits=0.3,
      ymin =0,
      ymax = 17,
      xtick = data,
      symbolic x coords= {$D_{16}$, $D_{76}$, $D_{42}$}, 
      ylabel=Loadings in \%,
      bar width = 0.5cm,
      ] 
      \addplot [draw=blue,fill=blue, opacity=0.5] 
      coordinates {($D_{16}$, 13.01) ($D_{76}$, 11.08) ($D_{42}$, 9.41)};
      
      \addplot[ dashed, black, sharp plot, update limits=true] 
      coordinates {($D_{16}$, 4.16) ($D_{42}$, 4.16)}; 
      \end{axis} 
 \end{tikzpicture}

%% file: plots/growthSchoolA.tex
\pgfplotsset{
tick label style={font=\scriptsize}, 
compat=newest, 
width=17cm, height=5cm, 
label style={font=\scriptsize}, 
}
\begin{tikzpicture}
      \begin{axis}[ybar, 
      enlarge x limits=0.05,
      enlarge y limits=0.10,
      xtick = data,
      bar width = 0.11cm,
      symbolic x coords= {$D_{16}$, $D_{19}$, $D_{20}$, $D_{24}$, $D_{28}$, $D_{40}$, $D_{42}$, $D_{49}$, $D_{50}$, $D_{51}$, $D_{52}$, $D_{53}$, $D_{54}$, $D_{55}$, $D_{56}$, $D_{57}$, $D_{58}$, $D_{64}$, $D_{65}$, $D_{67}$, $D_{71}$, $D_{72}$, $D_{76}$, $D_{78}$}, 
      ylabel=correct answers in \%,
      xlabel = Descriptors,
      legend style={at={(0.15, 0.9)},
		anchor=north,legend columns=-1},
      ] 
      \addplot [draw=blue,fill=blue, opacity=0.5] 
      coordinates {($D_{16}$, 17.11) ($D_{19}$, 31.08) ($D_{20}$, 23.73) ($D_{24}$, 17.24) ($D_{28}$, 27.59) ($D_{40}$, 19.30) ($D_{42}$, 28.57) ($D_{49}$, 37.11) ($D_{50}$, 21.88) ($D_{51}$, 27.63) ($D_{52}$, 55.79) ($D_{53}$, 19.35) ($D_{54}$, 15.00) ($D_{55}$, 15.79) ($D_{56}$, 28.81) ($D_{57}$, 36.36) ($D_{58}$, 22.37) ($D_{64}$, 28.07) ($D_{65}$, 21.05) ($D_{67}$, 14.04) ($D_{71}$, 28.21) ($D_{72}$, 23.68) ($D_{76}$, 81.03) ($D_{78}$, 32.89)};
      
      \addplot [draw=green,fill=green, opacity=0.5] 
      coordinates {($D_{16}$, 34.48) ($D_{19}$, 27.59) ($D_{20}$, 25.00) ($D_{24}$, 42.19) ($D_{28}$, 30.23) ($D_{40}$, 21.84) ($D_{42}$, 44.32) ($D_{49}$, 28.74) ($D_{50}$, 31.82) ($D_{51}$, 34.48) ($D_{52}$, 47.13) ($D_{53}$, 44.83) ($D_{54}$, 27.69) ($D_{55}$, 32.94) ($D_{56}$, 40.00) ($D_{57}$, 68.24) ($D_{58}$, 40.91) ($D_{64}$, 16.47) ($D_{65}$, 28.41) ($D_{67}$, 15.73) ($D_{71}$, 46.88) ($D_{72}$, 35.63) ($D_{76}$, 78.16) ($D_{78}$, 14.94)};
      
      \addplot [draw=orange,fill=orange, opacity=0.5] 
      coordinates {($D_{16}$, 53.47) ($D_{19}$, 39.24) ($D_{20}$, 32.99) ($D_{24}$, 43.42) ($D_{28}$, 61.17) ($D_{40}$, 51.96) ($D_{42}$, 60.00) ($D_{49}$, 35.64) ($D_{50}$, 62.89) ($D_{51}$, 51.96) ($D_{52}$, 54.64) ($D_{53}$, 52.94) ($D_{54}$, 27.63) ($D_{55}$, 40.20) ($D_{56}$, 56.00) ($D_{57}$, 66.67) ($D_{58}$, 58.76) ($D_{64}$, 43.59) ($D_{65}$, 33.66) ($D_{67}$, 38.38) ($D_{71}$, 30.67) ($D_{72}$, 53.00) ($D_{76}$, 76.77) ($D_{78}$, 39.22)};
      
      \legend{2016,2017,2018}

\end{axis} 
\end{tikzpicture}

%% file: discussion.tex
\section{Discussion}
\label{sec:discussion}

In this paper we presented the LPCA~\cite{2015arXiv151006112L} as a visualization tool for the analysis of math assessments. We applied the LPCA with the SPAECE math assessments from 2016 to 2018, in order to understand possible intrinsic relation among descriptors that have not been noiced so far. Then, using the proposed color maps, we investigated the distribution of the math abilities  of the population in more details.

Our findings have  been used to guide new educational policies of the Secretariat of Education for the State of Cear\'a (SEDUC-CE).  In particular, the formulation of a new curriculum taking into account the deficiencies in number sense of the students in Cear\'a is underway in the educational system.

Although the examples of this paper are limited to large-scale assessment in
Mathematics, it is still possible to reproduce them for other areas. However, we alert that, depending on the context of the area and the distribution of the abilities of the population, the interpretation of the visualizations may not follow  similar conclusions as those found in Section~\ref{sec:applications}. 

%% file: acknowledments.tex
\section*{Acknowledgments}

The authors are grateful to  SEDUC-CE for providing the data we uses in this
paper, especially to Alesandra Benevides(UFC-Sobral and EducLab) and George
Gomes (SEDUC-CE) for helping us to pre-process it. Thanks to Thomaz
Veloso (Insper and UFC) who helped  with proofreading. We acknowledge Funda\c{c}\~ao Cearense de Apoio ao Desenvolvimento Cient\'ifico e Tecnol\'ogico (FUNCAP) for the valuable financial support.

%% file: normalDeviance.tex
\section{Normal deviance}
\label{ap:normalDeviance}

The Gaussian probability density function with parameter $\phi$ is 
\begin{equation*}
f(x, \phi) = \frac{1}{\sqrt{2\pi}} \exp{\left\{-\frac{(x-\phi)^2}{2}\right\}}.
\end{equation*}

\noindent Now, observe that
$$l(\Theta, X) = \sum_{i,j}\left\{ -\frac{1}{2} \log(2\pi) - \frac{(x_{ij}-\theta_{ij})^2}{2}\right\}$$

\noindent and,

$$l(X, X) = \sum_{i,j}-\frac{1}{2} \log(2\pi).$$ 

\noindent Therefore, the scaled deviance is 
\begin{align*}
D(X, \Theta)  & = 2l(X, X)-2l(\Theta, X) =  
\sum_{i, j} (x_{ij} - \theta_{ij})^2 = \sum_{i} ||\bs{x_i} - \bs{\theta_i}||^2.
\end{align*}

%% file: bernoulliDiviance.tex
\section{Bernoulli deviance}
\label{ap:bernoulliDeviance}

The Bernoulli probability mass  function with parameter $p$ is 
given by
\begin{equation*}
f(x \;;p) = \left\{
\begin{array}{rl}
p & \text{if } x = 1,\\
1-p & \text{if } x = 0,\\
\end{array} \right.
\end{equation*}

It is also convenient to  express such function as $f(x \;;p)= p^x(1-p)^{1-x}$
\noindent. Now, using that 

$$\theta_{ij} = \displaystyle \textrm{logit} \; p_{ij} = \frac{p_{ij}}{1-p_{ij}} \iff p_{ij} = \frac{\exp{\theta_{ij}}}{1+ \exp{\theta_{ij}}}$$ .

\noindent we have,

\begin{align*}
\log f(x_{ij}\; ;  p_{ij}) &  = \log \{p_{ij}^{x_{ij}} (1-p_{ij})^{1-x_{ij}}\}  =  x_{ij}\log{p_{ij}} +(1-x_{ij}) \log{(1-p_{ij})} = \\
             & = x_{ij} \log\left(\frac{p_{ij}}{1-p_{ij}}\right) + \log{(1-p_{ij})} = x_{ij}\theta_{ij} + \log{\left(1- \frac{\exp \theta_{ij}}{1+\exp{\theta_{ij}}} \right)} = \\
             & = x_{ij}\theta_{ij} + \log{\left(\frac{\exp \theta_{ij}+1 -\exp \theta_{ij} }{1+\exp{\theta_{ij}}} \right)} = x_{ij}\theta_{ij} + \log{(1+ \exp{\theta_{ij}})^{-1}} = \\
             & = x_{ij}\theta_{ij} - \log{(1+ \exp{\theta_{ij}})}
\end{align*}

\noindent and it is easy to see that, $\log f(x_{ij}\; ; x_{ij})  = 0$. Thus, the Bernoulli scaled deviance is 

\begin{align*}
D(X, \Theta)  & = 2l(X, X)-2l(\Theta, X) = 2\sum_{i,j} \log f(x_{ij}\;;
    x_{ij}) -2\sum_{i, j}\log f(x_{ij}\; ;\theta_{ij}) = \\
              & = -2tr(X\Theta^T) + 2\sum_{ij} \log (1+\exp\,\theta_{ij}).
\end{align*}

%% file: referenceMatrix.tex
\section{SPAECE's Reference Matrix}
\label{ap:referenceMatrix}

The SPAECE's reference matrix is formed by a set of minimum expected skills (descriptors) in their various levels of complexity, in each area of knowledge and each stage of schooling. The matrices are based on studies of the curricular proposals of teaching in the current curricula of the Brazil, besides researches in didactic books and debates with active educators and specialists in education. The reference matrices are elaborated without the pretension of exhausting the repertoire of the necessary skills to the full development of the student. Therefore, they should not be understood as unique skills to be worked on in the classroom. Its purpose is to mark the creation of test items, which distinguishes them from curricular proposals, teaching strategies and pedagogical guidelines. Below, we describe the Reference Matrix of the third year of high school, which is the main focus of this paper.

\begin{table}
\centering
\begin{tabular}{|c|p{12cm}|}
\hline
\rowcolor{Gray}
\multicolumn{2}{|c|}{THEME I: INTERACTING WITH NUMBERS AND FUNCTIONS}\\
\hline
$D_{16}$    &  Establish relations between fractional and decimal representations of rational numbers.  \\
\hline
$D_{19}$    &  Solve problems involving simple interests. \\
\hline
$D_{20}$    &  Solve problems involving compound interests. \\
\hline
$D_{24}$    &  Factor and simplify algebraic expressions. \\
\hline
$D_{28}$    &  Identify the algebraic representation or graph of a polynomial function of 1st degree. \\
\hline
$D_{40}$    &  Relate the roots of a polynomial with its decomposition in factors of 1st degree. \\
\hline
$D_{42}$    &  Recognize algebraic or graphical representation  polynomial function of the 1st degree.\\
\rowcolor{Gray}
\hline
\multicolumn{2}{|c|}{THEME II: LIVING WITH GEOMETRY}\\
\hline
$D_{49}$    &  Solve problems involving similarities of planar figures.\\
\hline
$D_{50}$    &  Solve situation problem involving Pythagorean  Theorem or other metric relations in the right triangle. \\
\hline
$D_{51}$    &  Solve a problem using polygon properties (sum of internal angles, number of diagonals, computing the interior angle regular polygons).\\
\hline
$D_{52}$    &  Identify flattening  of some polyhedral and/or round objects. \\
\hline
$D_{53}$    &  Solve situation problem involving trigonometric ratios in right triangles (sine, cosine and tangent).\\
\hline
$D_{54}$    &  Calculate the area of a triangle given the coordinates of the vertices.\\
\hline
$D_{55}$    &  Determine the equation of a straight line given two points or a point and its inclination. \\
\hline
$D_{56}$    &  Among the equations of 2nd degree with two unknowns identify, those that represent a circumference.\\
\hline
$D_{57}$    &  Find the location of points in the Cartesian plane. \\
\hline
$D_{58}$    &  Interpret geometrically the coefficients of a straight line equation. \\
\rowcolor{Gray}
\hline
\multicolumn{2}{|c|}{THEME III: LIVING THE MEASURES}\\
\hline
$D_{64}$    &  Solve a problem using the relations between different measure unities of capacity and volume. \\
\hline
$D_{65}$    & Calculate the perimeter of planar figures in a situation problem.\\
\hline
$D_{67}$    & Solve problem involving calculation of areas of planar figures. \\
\hline
$D_{71}$    & Calculate the total surface area of prisms, pyramids, cones, cylinders and spheres.\\
\hline
$D_{72}$    & Calculate the volume of of prisms, pyramids, cones, cylinders in a situation problem.\\
\rowcolor{Gray}
\hline
\multicolumn{2}{|c|}{THEME IV: TREATMENT OF INFORMATION }\\
\hline
$D_{76}$    & Assign information presented in lists and/or tables to the graphs that represent them, and vice versa.\\
\hline
$D_{78}$    & Solve problem involving central tendency measures: mean, mode and median.\\
\hline
\end{tabular}
\caption{Reference Matrix for Middle School.}
\end{table}